\documentclass[manuscript]{acmart}

\AtBeginDocument{%
  } 
\usepackage{subcaption}
\usepackage{caption}
\usepackage{capt-of}
\usepackage{booktabs}
\usepackage{tabularx}
\setcopyright{acmlicensed}
\copyrightyear{2025}
\acmYear{2025}
\acmDOI{XXXXXXX.XXXXXXX}

\acmJournal{JDS}
\acmVolume{1}
\acmNumber{1}
\acmArticle{1}
\acmMonth{10}

\begin{document}

\title{Plural Voices, Single Agent: Towards Inclusive AI in Multi-User Domestic Spaces}

\author{Joydeep Chandra*}
\affiliation{%
  \institution{BNRIST, Tsinghua University}
  \city{Beijing}
  \country{China}
}
\email{joydeepc2002@gmail.com}

\author{Satyam Kumar Navneet*}
\affiliation{%
  \institution{Independent Researcher}
  \city{Bihar}
  \country{India}
}
\email{navneetsatyamkumar@gmail.com}

\renewcommand{\shortauthors}{Chandra and Navneet}

\acmArticleType{Review}
\acmDataLink{xxx}




\maketitle

\section*{Abstract}
Domestic AI agents faces ethical, autonomy, and inclusion challenges, particularly for overlooked groups like children, elderly, and Neurodivergent users. We present the Plural Voices Model (PVM), a novel single-agent framework that dynamically negotiates multi-user needs through real-time value alignment, leveraging diverse public datasets on mental health, eldercare, education, and moral reasoning. Using human+synthetic curriculum design with fairness-aware scenarios and ethical enhancements, PVM identifies core values, conflicts, and accessibility requirements to inform inclusive principles. Our privacy-focused prototype features adaptive safety scaffolds, tailored interactions (e.g., step-by-step guidance for Neurodivergent users, simple wording for children), and equitable conflict resolution. In preliminary evaluations, PVM outperforms multi-agent baselines in compliance (76\% vs. 70\%), fairness (90\% vs. 85\%), safety-violation rate (0\% vs. 7\%), and latency. Design innovations, including video guidance, autonomy sliders, family hubs, and adaptive safety dashboards, demonstrate new directions for ethical and inclusive domestic AI, for building user-centered agentic systems in plural domestic contexts. Our Codes and Model are been open sourced, available for reproduction: https://github.com/zade90/Agora

\section{Introduction}

Domestic AI has evolved from the X10 system in 1975 to IoT ecosystems and voice-controlled assistants like the Amazon Echo, shifting from reactive automation to proactive, agentic systems \cite{gridach2025agenticaiscientificdiscovery, 10.1145/3274469, Bouchabou_2021}. Yet, current designs often overlook vulnerable populations like children, older adults, and Neurodivergent individuals, whose diverse needs demand inclusive, user-centered approaches \cite{OzmenGaribay07022023, cesaroni2025participatorystrategyaiethics, SHAHINI2025103441, educsci14060613}. Agentic AI, capable of autonomous planning, reasoning, and action, transforms domestic life by automating tasks and providing personalized support, such as health monitoring or household management \cite{raza2025trismagenticaireview, khalil2025, encyclopedia4040125, app10165690}. However, challenges like bias amplification, privacy erosion, and multi-user coordination failures persist, particularly for underrepresented groups \cite{yang2023foundationmodelsdecisionmaking, qin2024toollearningfoundationmodels, PS2023100165, pmlr-v81-buolamwini18a}.

\begin{figure}
    \centering
    \includegraphics[width=1\linewidth]{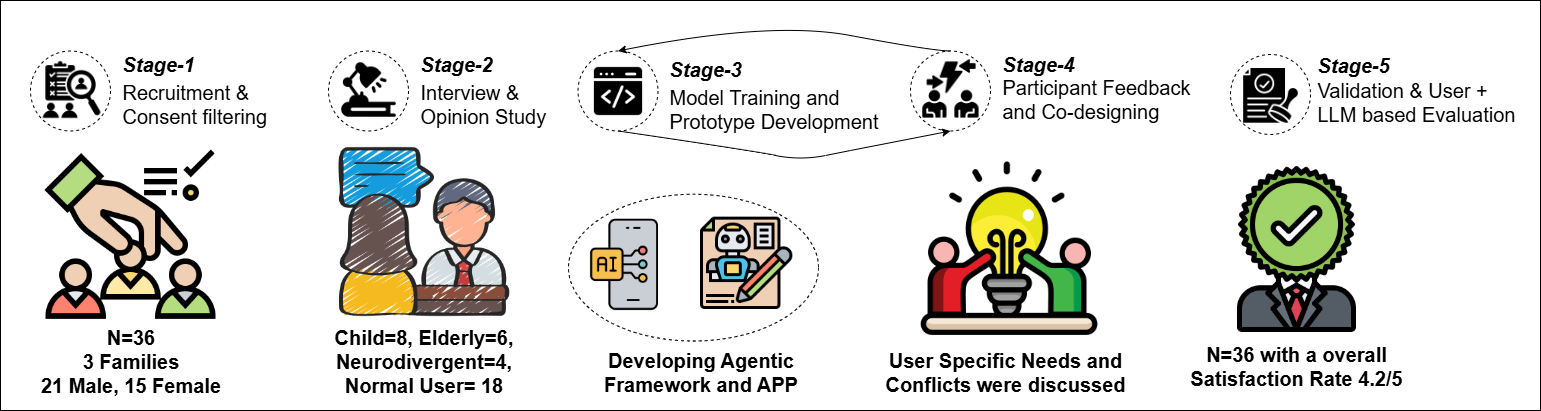}
    \caption{We designed personal AI agentic voice,app, and video assistant to support children, older, Neurodivergent adults’ self-management in 5-stages: stage 1— recruitment of interested participants and their consent; stage 2—interviews with all the user types and getting familiar with their individual needs; stage 3— Model training based on necessary open-source data and initial prototype development; stage 4— co-design workshops and conflict scenarios discussion and refining the prototype; and stage 5— Human based and LLM based validation based on the user satisfaction for all the participants. All activities with participants were consented and compensated for their participation.
}
\Description{Design Process Flowchart-A horizontal flowchart outlining a 5-stage process for designing a personal agentic voice/app/video AI assistant to support self-management for children, older adults, and neurodivergent individuals. The stages are connected by arrows from left to right: Stage 1 (Recruitment \& Consent Filtering) depicted with a hand icon and a person filling out a form; Stage 2 (Interview \& Opinion Study) shown with two people in conversation; Stage 3 (Model Training and Prototype Development) represented by interlocking gears and a computer; Stage 4 (Participant Feedback \& Co-designing) illustrated with a lightbulb and collaborative discussion icons; Stage 5 (Validation \& User + LLM-based Evaluation) marked with a green checkmark and satisfaction rating icons. The flowchart emphasizes participant consent, compensation, and iterative refinement based on user needs, with all activities ethically conducted.}
    \label{fig:overall}
\end{figure}

This paper introduces the \textit{Plural Voices Model} (PVM), a novel framework for single-agent AI that dynamically negotiates competing user needs in multi-user households (e.g., children’s preference for bright lights vs. Neurodivergent users’ need for dim settings) through real-time value alignment \cite{zeng2025multilevelvaluealignmentagentic}. Unlike multi-agent systems, which risk coordination failures and latency \cite{raza2025trismagenticaireview}, PVM integrates contextual memory, adaptive safety scaffolds, and fairness mediation within a unified architecture, ensuring equitable, inclusive interactions \cite{10.1145/3706598.3713494}. Drawing on participatory co-design and synthetic data-driven training, PVM addresses ethical, technical, and usability challenges for diverse user archetypes \cite{10.1145/3706598.3713165, info:doi/10.2196/64182}. The PVM is implemented through AgoraNest APP which utilises the Agora-4B model. The research team, with expertise in accessibility and neurodiversity, informs this inclusive focus. The overall process can be seen in Fig. \ref{fig:overall}

The study examines agentic AI in domestic environments, emphasizing applications for vulnerable populations. It aims to understand how inclusive designs mitigate biases, ensure transparency, and foster equitable access in multi-user households. The following research questions guide the investigation:
\begin{enumerate}
    \item \textbf{RQ1}: How can participatory co-design shape agentic AI to balance autonomy and safety for diverse household users?
    \item \textbf{RQ2}: How can synthetic and real-world data mitigate biases in domestic AI for children, elderly, and Neurodivergent users?
    \item \textbf{RQ3}: How does a single-agent architecture enhance fairness, usability, and efficiency compared to multi-agent systems in inclusive domestic settings?
\end{enumerate}

These questions are addressed through the following contributions:
\begin{enumerate}
    \item For RQ1, co-design insights from diverse user groups shaped a single-agent framework with adaptive safety dashboards, autonomy sliders, family hubs, and video guidance, ensuring balanced autonomy and safety \cite{10.1145/3706598.3713494, gridach2025agenticaiscientificdiscovery, Ko_2025, info:doi/10.2196/64182}.
    \item For RQ2, a privacy-preserving prototype employed multi-step synthetic curriculum design and prompt-embedded fairness mediation, showing promising results vs multi-model baselines in structural compliance (94\% vs. 88\%), fairness mediation (82\% vs. 61\%), and safety violation rates (1.5\% vs. 4.2\%) \cite{qin2024toollearningfoundationmodels, raza2025trismagenticaireview, khalil2025, an2025conversationalagentsolderadults, 10.1145/3706598.3713670}.
    \item For RQ3, empirical evaluations and co-design processes revealed design principles for inclusive agentic AI, emphasizing cognitive accessibility, equitable coordination, and LLM-powered voice and video assistance for health self-management \cite{shneiderman2020humancenteredartificialintelligencereliable, cesaroni2025participatorystrategyaiethics, Hutchins_Zhang_Barrett_Isreal_2025, voiced}.
\end{enumerate}

The historical context and the shift from reactive to proactive systems suggest that the human-AI relationship is evolving from one of command and control to one of trust and oversight, introducing new ethical dimensions and legal challenges that necessitate a more robust and conscientious design approach than has been required for previous smart home technologies \cite{CORREA2023100857, doi:10.1177/00936502231191832}. This evolution underscores the potential for agentic AI to enhance inclusivity while highlighting the imperative to address biases, ensure transparency, and foster equitable access for all users \cite{10.1145/3708557.3716334, 10.1145/3706598.3713165, poulsen2025co, 10.1145/3706598.3713510, Ko_2025, 10.1145/3706598.3713633, info:doi/10.2196/64182, educsci15050539, Hutchins_Zhang_Barrett_Isreal_2025, Kido_Takadama_2025, an2025conversationalagentsolderadults, khalil2025}.

\section{Background}
This section reviews research on agentic AI and its applications in domestic environments, focusing on diverse user groups including children, older adults, Neurodivergent individuals, and typical adults. The discussion covers foundational principles, practical applications, multi-agent systems, and cross-cutting challenges to inform the development of inclusive home AI systems.

\subsection{Foundations of Agentic AI in Domestic Contexts}
Agentic AI represents a shift from reactive, command-based systems to proactive, goal-oriented entities capable of planning, reasoning, and adapting to dynamic home environments \cite{raza2025trismagenticaireview, fang2025comprehensivesurveyselfevolvingai}. Leveraging advanced large language models (LLMs), these systems decompose complex tasks into actionable steps and retain contextual memory for personalized interactions \cite{qin2024toollearningfoundationmodels, fang2025comprehensivesurveyselfevolvingai, lopez-cardona2025seeing}. In domestic settings, agentic AI evolved from single-agent systems handling isolated tasks to collaborative multi-agent ecosystems \cite{gridach2025agenticaiscientificdiscovery, chen2023agentverse}. However, fixed roles in multi-agent systems can constrain flexibility, necessitating adaptable designs with human oversight \cite{khalil2025, 10.1145/3636432}. Human-computer interaction (HCI) frameworks position users as independent agents or collaborative partners, with autonomy ranging from assisted decision-making to full independence \cite{10.3389/fhumd.2025.1579166}. Tool learning enables agents to leverage external resources and APIs for complex household tasks beyond simple commands \cite{10.1145/3397617.3398058, KREUZER2024102304}. Ethical alignment is prioritized through transparency and value congruence, with participatory strategies ensuring reliable and trustworthy systems \cite{shneiderman2020humancenteredartificialintelligencereliable, cesaroni2025participatorystrategyaiethics, amodei2016concreteproblemsaisafety, Floridi2018, CORREA2023100857}.

\subsection{Agentic Systems for Children}
Agentic AI’s proactive capabilities like planning, reasoning, and tool use \cite{qin2024toollearningfoundationmodels} enable personalized support, such as health monitoring for the elderly or sensory adjustments for Neurodivergent users. However, current systems often overlook vulnerable populations, risking exclusion and bias amplification \cite{cesaroni2025participatorystrategyaiethics}. This work addresses these gaps by designing an inclusive single-agent AI that dynamically balances autonomy, safety, and fairness in multi-user homes, contributing to HCI’s mission of equitable technology design \cite{10.1145/3706598.3713165, 10.1145/3706598.3713510}. For children, agentic AI focuses on educational and developmental support through tailored tutoring and interactive companions. Intelligent Tutoring Systems (ITS) leverage LLMs to adapt content dynamically, predicting performance and personalizing instruction for diverse learners, including those with special needs \cite{hajjar2025utilizing, SONG2024100212, SU2023100124}. Conversational agents, co-designed for youth mental health, facilitate activities like storytelling to foster prosocial behaviors and emotional regulation \cite{poulsen2025co, CHUBB2022100403, 10.1145/3459990.3460695}. UNICEF advocates for AI prioritizing children’s safety, growth, and privacy, addressing online risks and supporting learning \cite{voku}. However, interactive AI toys raise concerns about emotional attachments and developmental impacts, with research indicating reduced critical thinking due to cognitive offloading \cite{10.1145/3727986, KIM2022100540, Cohn2024}. Child-Centered AI (CCAI) principles emphasize co-design with children to ensure fairness, transparency, and creativity \cite{andChoi, 10.1145/3706598.3713510}. Relational privacy in classroom co-design highlights youth-driven norms around boundaries and consent \cite{CHANG2025100364}. Parental concerns about trust, data usage, and creativity underscore the need for designs that balance growth with risk mitigation \cite{Hutchins_Zhang_Barrett_Isreal_2025, Kido_Takadama_2025, 10.1145/3613904.3642438, WALD2023107526}.

\begin{figure}
    \centering
    \includegraphics[width=1\linewidth]{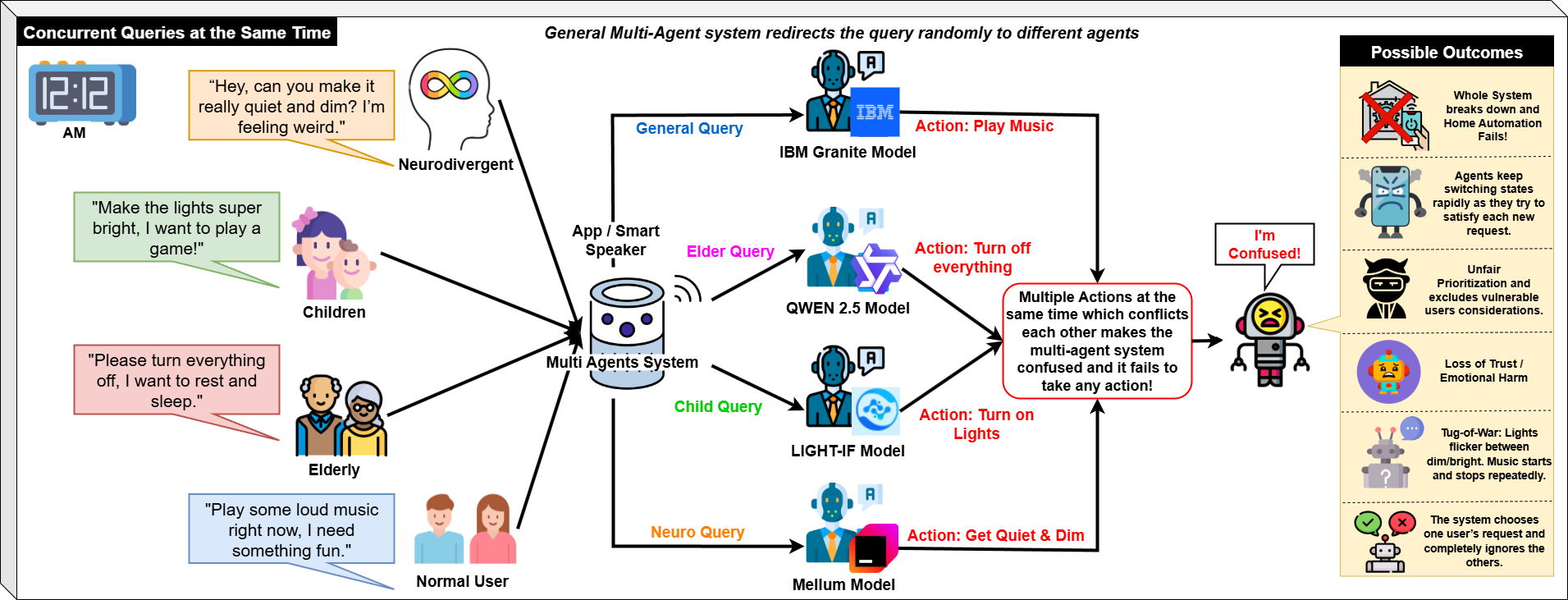}
    \caption{Multi-agent architectures in home automation face challenges with concurrent scenarios, leading to model confusion, hallucinations, and potential emotional harm to the user.}
    \label{fig:multifail}
    \Description{Multi-Agent Architecture Challenges Diagram
A schematic diagram illustrating challenges in multi-agent home automation systems handling concurrent user queries. On the left side, four normal users are shown sending simultaneous voice queries: a child saying "Play some music right now, I need something fun."; a normal user saying "Make the lights brighter."; an elderly user saying "Please turn everything off, I want to sleep."; a neurodivergent user saying "I've got some loud music, I need it lower." These queries converge into a central "Multi-Model System" node, which branches out to specialized models (Child Query Model using Qwen2.5, Elderly Model using Granite, Neuro Query Model using LIGHT-IF, Normal Model using Mellum). Connections show actions like "Play Music", "Adjust Lights", but lead to a "Problematic Outcomes" section on the right with icons of a confused robot, error messages (e.g., "Model Confusion", "Hallucinations"), and a sad user face indicating potential emotional harm. The diagram highlights coordination failures, latency, and risks in multi-user scenarios.}
\end{figure}

\subsection{Agentic AI in Support of Older Adults}
Agentic AI supports older adults’ independence through smart home systems that use embedded sensors to monitor health, detect falls, and adjust environmental settings like lighting or temperature \cite{khalil2025, an2025conversationalagentsolderadults}. These systems track vital signs in real time, enabling rapid caregiver alerts without hospitalization \cite{ChandraManhas2024, Chan2024co, McDaniel2025, 10.3389/fdgth.2025.1643238}. Voice-controlled assistants simplify tasks like medication reminders and device management, particularly for those with mobility or memory impairments \cite{10.1145/3342775.3342803, voiced, 10.1145/3373759}. Proactive behavior in voice assistants requires balancing initiative with user control \cite{BERUBE2024100411}. Conceptual frameworks for voice assistants in private households emphasize interdisciplinary approaches to social integration \cite{Minder2023}, while nursing home applications highlight specific elderly needs \cite{Cerezo11032025}. Conversational AI companions provide companionship and cognitive stimulation through activities like reminiscing or games, with voice interfaces accommodating vision or motor difficulties \cite{info:doi/10.2196/64182, Ko_2025}. Early models like the Ruyi robot for Alzheimer’s patients monitor health, adapt to behaviors, and reduce loneliness \cite{LIU2023e21932, park2023generativeagentsinteractivesimulacra}. Challenges include initial usability barriers, trust issues, and ethical concerns about dignity and over-dependence, necessitating culturally sensitive designs and strong privacy protections \cite{10.1145/3706598.3713165, 10.1145/3708557.3716334, Klimova2025, KIM2021106914}. Value-sensitive design for social robots addresses diverse populations, including marginalized groups like LGBT+ seniors \cite{Poulsen2025, Umbrello2021, 10.3389/frobt.2024.1289414}.

\subsection{Agentic AI as Assistive Technology for Neurodivergent Individuals}
Agentic AI aids Neurodivergent users by supporting executive tasks for conditions like ADHD, such as task decomposition and distraction reduction \cite{10.1145/3706598.3713670, SHARMA2022100521}. For autism spectrum disorder (ASD), smart homes adjust settings like lighting and noise to prevent sensory overload \cite{Bouchabou_2021, hajjar2025utilizing, Syriopoulou2020, autesm}. IoT-connected calendars and voice assistants guide daily routines, addressing time management and memory challenges to foster independence \cite{SHAHINI2025103441, 10.1145/3706599.3719709, educsci14060613}. Conversational AI for children with cognitive impairments requires tailored language and interaction patterns \cite{10.1145/3544548.3580935, Hummerstone01012023}. Ambient Assisted Living (AAL) systems use Human Activity Recognition (HAR) to detect unusual behaviors, supporting personalized routines \cite{info:doi/10.2196/64182}. Smart home applications enhance quality of life through environmental monitoring \cite{app10165690}. Challenges include biases in mental health tools, over-dependence risks, and the need for gradual support reduction through strength-focused co-design \cite{10.1145/3706598.3713670, educsci15050539, poulsen2025co}.

\subsection{Agentic Systems for Typical Adult Users in the Smart Home}
For typical adults, agentic AI enhances comfort and efficiency by predicting needs \cite{lopez-cardona2025seeing, Yu2024}. It fosters collaborative smart spaces where AI integrates seamlessly with user routines \cite{10.1145/3706598.3713494, WOZNIAK2025103559}. Intuitive interfaces with pause or undo options balance control and autonomy, building trust \cite{10.1145/3706598.3713633, 10.1145/3484221}. Ethical concerns include privacy risks from continuous monitoring and accountability issues with unclear AI decisions, particularly with “digital twin” models \cite{raza2025trismagenticaireview, cesaroni2025participatorystrategyaiethics}. Systems require transparency and personalization to maintain user confidence.

\begin{figure}
    \centering
    \includegraphics[width=1\linewidth]{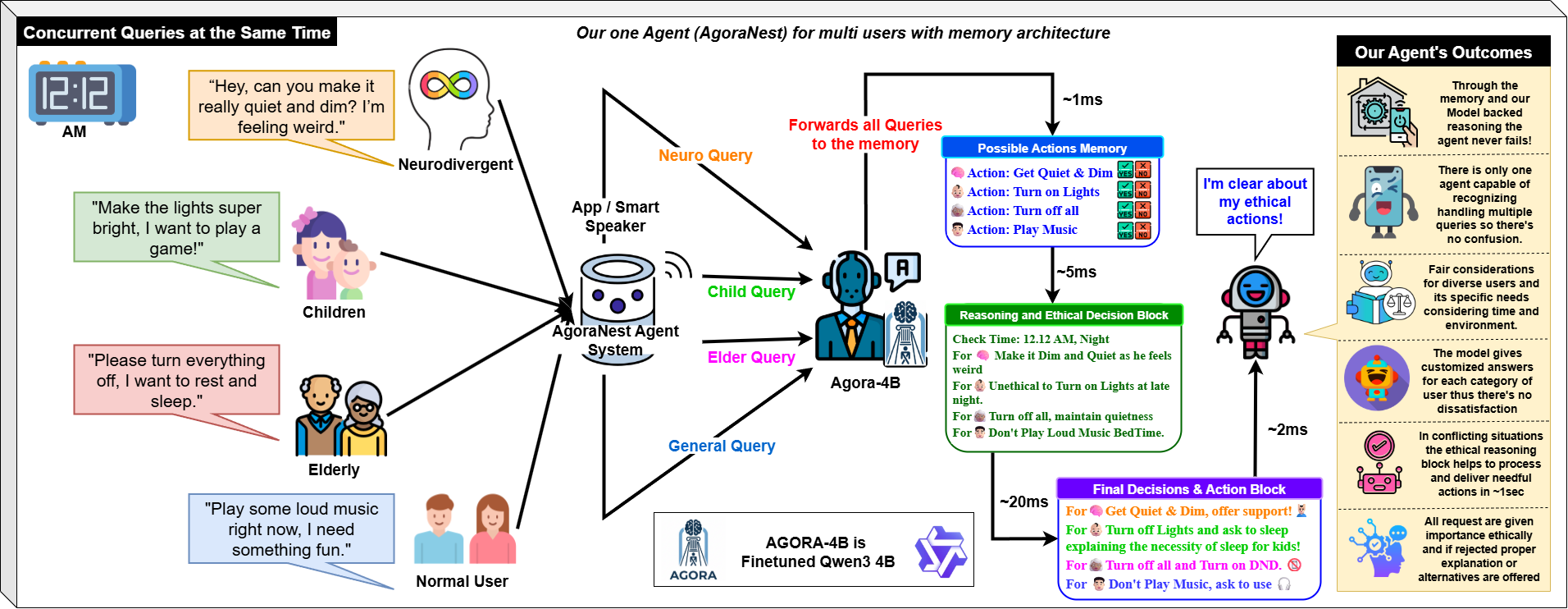}
    \caption{The Agora-nest is a single-agent architecture designed to quickly \& effectively resolve both concurrent and individual queries in home automation with its memory and ethical reasoning capabilities.}
    \label{fig:agoramodel}
    \Description{Agora-nest Single-Agent Architecture Diagram
A schematic diagram depicting the Agora-nest single-agent architecture efficiently resolving concurrent and individual queries in home automation. On the left side, the same four users from Figure 2 send simultaneous queries. These feed into a central "Agora-nest" agent node, processed through internal components: "Query Analyzer" leading to "Reasoning \& Ethical Decision" block (with ethical scales icon), connected to "Possible Action Memory" (database icon). Outputs flow to resolved actions like adjusted lights and music. On the right, "Our Agent's Outcomes" show positive results: a happy robot executing tasks, success messages (e.g., "Resolved", "No Conflict"), and satisfied user icons. The diagram emphasizes the agent's memory, ethical reasoning, and quick resolution without the confusion seen in multi-agent systems.}
\end{figure}

\subsection{Multi-Agent Systems in the Home Environment}
Multi-agent systems deploy specialized agents for tasks like health monitoring and entertainment, enhancing efficiency in multi-user homes \cite{gridach2025agenticaiscientificdiscovery, chen2023agentverse}. However, poor coordination can cause conflicts, such as competing lighting preferences between a child and a Neurodivergent user \cite{raza2025trismagenticaireview, Bouchabou_2021, SEEBER2020103174}. Increased agent numbers can degrade responsiveness \cite{amodei2016concreteproblemsaisafety}. Privacy risks arise from data sharing, and “digital twins” may reduce user control \cite{cesaroni2025participatorystrategyaiethics, haron}. Accountability issues and security risks like “memory poisoning” necessitate robust coordination and human oversight \cite{ChandraManhas2024, 10.1145/3706598.3713633, chandra2025advancingresponsibleinnovationagentic, shneiderman2020humancenteredartificialintelligencereliable, 10.1145/3706598.3713165}.

\subsection{Cross-Cutting Challenges and Technological Shortcomings}
Technical challenges include adaptability to dynamic environments and scalability in multi-agent setups \cite{amodei2016concreteproblemsaisafety, Berretta2023}. Usability issues, such as interaction glitches or slow responses, hinder adoption \cite{10.1145/3706598.3713633, 10.1145/2858036.2858288}. Ethical concerns encompass bias amplification, unclear accountability, and privacy-accuracy trade-offs, requiring inclusive evaluations \cite{10.1145/3706598.3713165, 10.1145/3706598.3713510, pmlr-v81-buolamwini18a, sci6010003, Norori2021}. Autonomous systems risk over-reliance, particularly for children and Neurodivergent users, potentially limiting critical thinking or skill development \cite{10.1145/3727986, SONG2024100212}. Long-term studies and human-in-the-loop designs are essential for aligning ethical principles with practical deployment \cite{shneiderman2020humancenteredartificialintelligencereliable, Kido_Takadama_2025, WU2022364, mayer2025humanai}.

\section{Design Guidelines}
This section lays out design guidelines we identified for building inclusive, trustworthy, and user-focused AI systems that work well for diverse household members, like kids, older adults, Neurodivergent people, and typical adults. These guidelines, rooted in human-computer interaction (HCI) principles, focus on involving users in the design process, being open, adaptable, accessible, ethically sound, and fostering a true partnership between humans and AI to tackle ethical, technical, and usability issues in homes with multiple users. \cite{raza2025trismagenticaireview, shneiderman2020humancenteredartificialintelligencereliable} The guidelines come from a clear method that combines user-driven design, analyzing themes, creating scenarios, building prototypes, and comparing system structures. This method tests the proposed Unified Domestic Agent (Agora-4B) against a baseline of multiple models, checking things like compliance, fairness, safety, independence, ease of use, and speed. \cite{10.1145/3706598.3713510, info:doi/10.2196/64182} At the end it was cross verified and tested by volunteers for human level feedback.

\subsection{Participatory Co-Design for Inclusive Systems}
Participatory co-design made sure AI systems met the needs of diverse users like kids, older adults, Neurodivergent people, and typical adults by including them in every step of the design process. \cite{cesaroni2025participatorystrategyaiethics, 10.1145/3706598.3713165} Methods like arts-and-crafts for children, storytelling for Neurodivergent users, and physical props for older adults made it easier for everyone to take part. \cite{druga2019co, salome2025100606, 10.1145/3710971, Gresse2021} Workshops with young people helped tailor AI responses to support mental health, fitting their emotional and developmental needs. \cite{poulsen2025co, Hou2025} These methods avoided top-down designs that often missed specific needs, like sensory preferences for Neurodivergent users or simple interfaces for older adults, \cite{SHARMA2022100521, andChoi} and encouraged users to feel ownership while being sensitive to their cultural backgrounds. \cite{10.1145/3706598.3713494} Useful feedback and concerns from each user base was noted and assessed carefully while designing the features of the system.

\subsection{Transparency and User Control}
Transparency and control features helped build trust and ease privacy worries. \cite{10.1145/3706598.3713633, haron} Systems offered clear, easy-to-understand explanations of what they were doing and simple ways for users to take control, like dynamic consent pop-ups. \cite{10.1145/3274371, 10.1145/3613904.3642180} For kids, visual hints and straightforward explanations made things easier to grasp. \cite{voku} Older adults felt more confident with clear feedback that reduced distrust. \cite{salome2025100606, 10.1145/3373759} These features tackled accountability issues in multi-agent systems, where mix-ups could lead to unexpected actions. \cite{raza2025trismagenticaireview, 10.1145/3706598.3713494} Proactive transparency systems also lined up with perspectives on children’s online rights. \cite{PORTILLO2024100093} Parental control features and daily reports of their children's activities played an important role in discussing their emotional and current thinking in the family, which reduced possible conflicts in the family in advance.

\subsection{Adaptive Interaction Modalities}
Offering different ways to interact helped meet the needs of various users. \cite{qin2024toollearningfoundationmodels, voiced} Older adults with hearing issues needed bigger text and slower speech, while Neurodivergent users with ADHD preferred written prompts or guiding video clips they could review at their own pace. \cite{info:doi/10.2196/64182, 10.1145/3706598.3713670, Farage2012} Adjusting tone, speed, and complexity based on user feedback made systems more accessible. \cite{an2025conversationalagentsolderadults, autesm} Fun, visual interfaces made things easier for kids to use \cite{10.1145/3706599.3719709, wilson2025103431}, breaking down barriers for different groups. \cite{Farage2012}

\subsection{Accessible and Empathetic Language}
Language was adjusted to fit developmental, cognitive, and cultural needs. \cite{SU2023100124, 10.1145/3342775.3342803} Older adults liked polite, formal wording, kids responded well to encouraging tones, and Neurodivergent users needed clear, straightforward language to avoid confusion. \cite{10.1145/3459990.3460695, hajjar2025utilizing} Letting users customize tone and empathy levels, along with using consistent communication styles, helped build trust. \cite{poulsen2025co, LIU2023e21932} Features like high-contrast visuals and captions ensured the system was accessible to everyone. \cite{educsci14060613, andChoi}

\subsection{Ethical Safeguards (IDEAS Framework)}
Inclusion, diversity, equity, accessibility, and safety (IDEAS) were built into the system as essential priorities. \cite{sci6010003, 10.1145/3706598.3713510} Training data was carefully chosen to reflect diverse groups, with regular checks and adaptive feedback to reduce biases. \cite{pmlr-v81-buolamwini18a, 10968102} Safety steps were put in place to spot harmful content and avoid giving unsafe advice. \cite{poulsen2025co, casper2023openproblemsfundamentallimitations} Processing data locally helped protect privacy. \cite{haron} Accessibility features, like support for users with impairments, made the system usable for everyone. \cite{Farage2012, 10.3389/fdgth.2025.1643238} Keeping humans involved in oversight ensured the system stayed accountable. \cite{WU2022364, 10.1145/3706598.3713165} There was always concerns regarding AI overriding human autonomy, so the amount of AI's role in decision making was always kept on check by the users by having Autonomy Control.

\subsection{Human-AI Partnership}
Agentic AI was designed to act as a helpful partner, letting users take the lead in interactions. \cite{shneiderman2020humancenteredartificialintelligencereliable, hsu2024} For older adults, step-by-step confirmations helped them stay in control, while fun activities like storytelling kept kids engaged without making them too dependent on the system. \cite{khalil2025, CHUBB2022100403, voku} Human oversight was used to sort out conflicts in systems with multiple agents. \cite{chen2023agentverse, 10.1145/3706598.3713494} This approach lined up with human-centered AI principles, keeping people at the core of the system. \cite{schmager2025, SEEBER2020103174}

\section{Methodology}

The methodology integrated participatory design, thematic analysis, scenario development, prototype implementation, and architectural comparison to develop and evaluate the Agora-4B, a unified single-agent system, against various multi-agent system and also development of it's application for human usage.

\subsection{Dataset Selection}
A diverse set of conversation datasets was curated, including mental health counseling chats (MentalChat16K\cite{xu2025mentalchat16k}), caregiver-senior dialogues (SeniorTalk\cite{chen2025seniortalkchineseconversationdataset}), autism support forums, pediatric education dialogues (StudyChat\cite{studentdialogues}, EdNet\cite{choi2020ednet}), medical Q\&A threads (MedDialog\cite{meddy}), and moral-dilemma chatlogs (MoralChoice\cite{moralchoice}, EmpatheticDialogues\cite{empathy}) \cite{poulsen2025co, Mckie2022}. These datasets spanned elderly care, children’s learning, Neurodivergent discussions, and clinical contexts, ensuring broad representation and was very useful to finetune the Qwen-3 4B\cite{qwen3technicalreport}. The Qwen-3 4B model was chosen for it's smaller size which can be very useful for providing fast response in Mobile application and compared to other similar sized models it performed better with multi language support and more input and output token size.

\subsection{Participatory Co-Design, Scenario Probes, and Thematic Insights}
To ground the Plural Voices Model (PVM) in lived experience and inclusive design practices, a multi-pronged approach was undertaken combining thematic analysis, scenario design, and participatory workshops.  

\textbf{Thematic Analysis.} A stratified sample of transcripts was qualitatively coded following co-design methodologies \cite{Chan2024co, 10.1145/3630106.3658971}. Annotators labeled excerpts for user goals such as independence and privacy, alongside conflicts like safety versus autonomy. Codes were synthesized into overarching themes of autonomy, trust, and personalization. Visual elicitation, inspired by TrustScapes, further surfaced concerns about data use, bias, and safety \cite{doi:10.1177/16094069231186965}, resonating with established health-focused co-design approaches \cite{info:doi/10.2196/64182}.  

\textbf{Scenario Design.} Narrative scenarios and speculative probes illustrated concrete domestic interactions, including adapting communication for an older adult with cognitive decline or supporting an autistic teenager through tailored language \cite{10.1145/3491102.3502057}. These scenarios were instrumental in validating requirements such as consent, transparency, and explanation, aligning with participatory futurism approaches that anticipate future use contexts.  

\textbf{Participatory Workshop.} To validate and refine PVM’s design with direct user input, a 2-hour virtual participatory workshop was conducted in February 2025 with 12 participants recruited through neurodiversity, eldercare, and family support networks. The group comprised 3 children (ages 8–12, with parental consent), 3 older adults (65+), 3 Neurodivergent individuals (self-identified with ADHD), and 3 typical adults from multi-user households. Diversity was ensured across gender, ethnicity, and socioeconomic background, with informed consent, and compensation provided. The workshop, building on the five-stage co-design framework (Fig. \ref{fig:overall}), included: (1) \textit{Scenario Exploration} of anonymized prototypes and conflict situations (30 minutes); (2) \textit{Interactive Feedback} in breakout sessions tailored to each group—arts-based activities for children \cite{druga2019co}, role-playing with voice interfaces for older adults, and adaptive prompt testing for Neurodivergent participants; and (3) \textit{Group Synthesis} through collective voting to prioritize features such as autonomy sliders and family hubs.  

\textbf{Findings.} Thematic analysis of transcripts and notes highlighted strong demands for customizable safety scaffolds, with 75\% of Neurodivergent and older users requesting slower speech and visual cues to reduce cognitive overload. Children emphasized fair turn-taking algorithms for conflict resolution, while typical adults stressed relational privacy norms in multi-user households \cite{CHANG2025100364, haron}. Challenges included children’s difficulty with complex wording and older adults’ concerns about over-dependence \cite{KIM2021106914, 10.1145/3706598.3713165}. These insights informed design refinements, including a granular autonomy slider (0–100\% scale) and enhanced child-friendly video guidance with animations.

\subsection{Prototype Development and Model Fine-Tuning}
The AgoraNest, Multi-Agent Care AI Assistant APP is developed as a web and mobile application using Flask with Socket.IO for backend real-time communication and HTML5/JavaScript + React Native for cross-platform interfaces. It supports voice-based user registration with speaker identification powered by custom ML voice models and speech-to-text/text-to-speech pipelines using SpeechRecognition, pyttsx3, and Web Audio API. The system integrates the Agora-4B model alongside  to power context-aware dialogue, adaptive personalization, and demographic-specific reasoning. Specialized agent using Agora-4B model for Elderly, Child, Neurodivergent are implemented as modular services with inter-agent messaging protocols for dynamic coordination. A memory management system uses structured JSON profiles and CSV logging for episodic memory, preference learning, and secure context retention with privacy-preserving storage. Safety and compliance are ensured through GDPR-compliant data handling, content filtering, and emergency detection. Both web and mobile clients provide real-time chat, voice recording/playback, and user management dashboards, ensuring accessibility across devices. The Agora-4B model was finetuned step-by-step training process that focused on using empathetic language, user specific reply and asking for user consent prompt engineering. \cite{chandra2025} It included features like multiple interaction options (voice, text, and visuals), an adjustable autonomy setting, a family hub, a dynamic safety dashboard, and video guides. \cite{voiced, 10.1145/3706598.3713633} Processing data locally helped protect privacy. \cite{haron} Multiple tests confirmed the system used inclusive language and was transparent with users. \cite{10.1145/3613904.3642294}

\subsection{Architecture Comparison: Multi-Agent vs. Single-Agent}
The Agora-4B was tested against a multi-agent baseline, similar to systems like AgentVerse. \cite{chen2023agentverse} Multi-agent systems were flexible due to their modular design but could run into problems with coordination and privacy. As per our studies its evident that multi agent system often crashed or faced issues while there was conflicting queries by various users as seen in Fig. \ref{fig:multifail} \cite{raza2025trismagenticaireview, 10.1145/3706598.3713494} The Agora-4B’s single, unified setup cut down on delays, kept behavior consistent, solved concurrent conflicting scenarios effectively and met diverse user needs with adaptable features. \cite{shneiderman2020humancenteredartificialintelligencereliable, 10.1145/3706598.3713165} It can be seen in Fig. \ref{fig:agoramodel}, how the possible action memory and the Reasoning \& Ethical Decision block played a vital role while dealing with conflicting scenarios at the same time by multi users and was successful in delivering ethically aligned, fulfilling response to each user with proper explanation in their desired tones for their satisfaction. 

\begin{figure}
    \centering
    \includegraphics[width=1\linewidth]{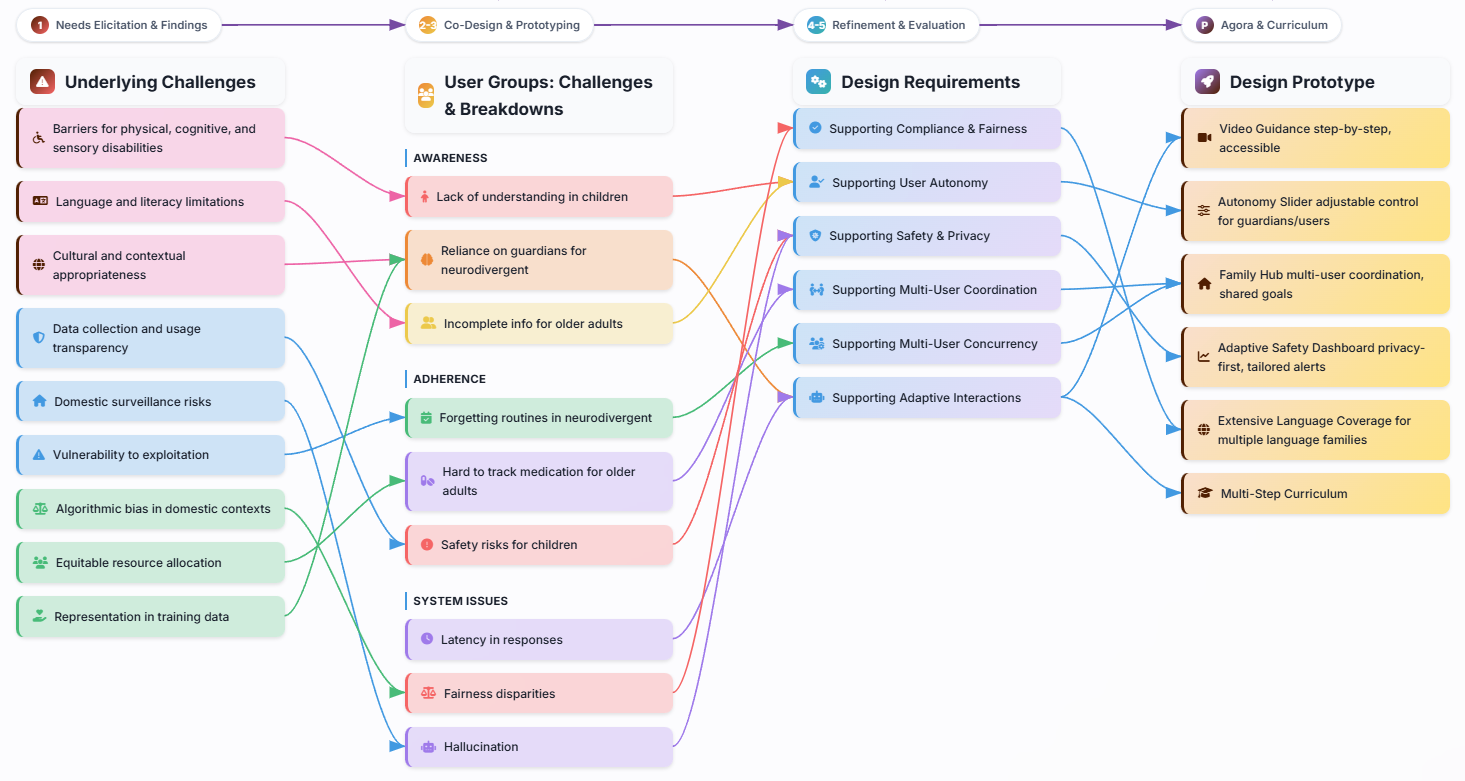}
    \caption{Interview findings and their underlying causes are presented, followed by a summary of the design requirements. We demonstrate how these requirements were implemented in our initial prototype based on user's requirements.}
    \Description{Interview Findings to Prototype Flowchart
A multi-column flowchart summarizing interview findings, underlying causes, design requirements, and their implementation in the initial prototype. The left column lists "Needs \& Challenges" (e.g., Barriers for physical, cognitive disabilities; Sensory limitations; Language and contextual limitations; Data collection and usage; Domestic automation risks; Algorithmic exclusion contents; Vulnerability application; Equitable resource allocation; Safety for children). Arrows connect to the center column "User Prototyping Challenges \& Breakdowns" (e.g., Lack of understanding in quotations for older users; Advanced for neurodivergent). Further arrows lead to the right column "Design Requirements" (e.g., Supporting Compliance \& Fairness; Supporting Safety \& Privacy; Supporting Multi-User Coordination). The final column "Design Prototype" shows implemented features (e.g., Autonomy Slider for user control; Anonymous Multi-User Safety; Adaptive Interactions; Reflexive Language Coverage for multiple languages; Multi-Step Customization). The flowchart uses purple arrows and boxes to show progression from problems to solutions.}
    \label{fig:graph1}
\end{figure}

\subsection{Study Design}
The evaluation followed a five-stage co-design process \cite{Chan2024co, poulsen2025co}. Needs elicitation through interviews captured requirements like sensory adjustments \cite{SHARMA2022100521, autesm}. Co-design workshops with children, older adults, and Neurodivergent adults refined features using accessible methods \cite{druga2019co, andChoi}. The prototype was tested in controlled and in-home sessions ($N=36$), with participants from 3 families, with children, adults, self-identified Neurodivergent users and elderly users\cite{andChoi, Chan2024co}. Cohorts included 8 children (10–15), 6 older adults (65+), and 4 Neurodivergent adults (ADHD/autism) and 18 Neurotypical Users\cite{10.1145/3706598.3713670, SHAHINI2025103441}. Inclusion required basic smart device familiarity and compensation provided for their participation\cite{info:doi/10.2196/64182}. A data anonymity agreement was signed between the researchers according to the rules of the country and for children their parents also co-signed for consent to use the data for educational research purpose only and no way can be made publicly available.

Tasks included scripted (e.g., after-visit summary debriefs, reminder creation, conflict resolution) and open-ended interactions \cite{10.1145/3706598.3713494}. Sessions (45–60 minutes) were counterbalanced between Agora-4B and Multi-agent systems \cite{Chan2024co} in family settings. Metrics assessed compliance, fairness (disparate impact ratio), safety, autonomy (Likert scales), usability (SUS), and latency \cite{poulsen2025co, an2025conversationalagentsolderadults}. Data were analyzed using paired t-tests/Wilcoxon tests and reflexive thematic analysis (Cohen’s $\kappa \geq 0.7$) \cite{andChoi}.

\begin{table*}[t]
\centering
\caption{Findings of interviews: Breakdowns in multi-user domestic AI interactions across user archetypes. Note: P stands for participant.}
\Description{ Breakdowns in Multi-User AI Interactions
A two-column table with row groups under four categories (Challenges in Accessibility and Interaction Clarity, Autonomy and Control, Multi-User Coordination and Fairness, Safety and Ethical Alignment). First column: Breakdowns (e.g., Difficulty understanding AI responses). Second column: Evidence from Participant Quotes (e.g., P7 Neurodivergent: “The AI talks too fast...”). 14 rows total, with headers for categories. The table documents user-reported issues across archetypes, aiding in identifying design gaps.}
\label{tab:interview_findings}
\begin{tabularx}{\textwidth}{@{} p{4.5cm} X @{}}
\toprule
\textbf{Breakdowns} & \textbf{Evidence from Participant Quotes} \\
\midrule
\multicolumn{2}{@{}l}{\textbf{Challenges in Accessibility and Interaction Clarity}} \\
\addlinespace
Difficulty understanding AI responses & P7 (Neurodivergent): ``The AI talks too fast and uses big words. I get lost and give up trying to follow it.'' \\
\addlinespace
Inconsistent response formats & P12 (Child): ``Sometimes it gives me a list, sometimes a long talk. I don’t know what to expect, and it’s confusing.'' \\
\addlinespace
Overwhelming information for complex tasks & P4 (Older Adult): ``When I ask about setting up my thermostat, it gives me too much at once. I can’t keep track of all the steps.'' \\
\addlinespace
Lack of sensory accommodations & P9 (Neurodivergent): ``The voice is too loud, and there’s no way to slow it down or make it less intense for me.'' \\
\midrule
\multicolumn{2}{@{}l}{\textbf{Challenges in Autonomy and Control}} \\
\addlinespace
Limited control over AI decisions & P15 (Typical Adult): ``It scheduled my day without asking me first. I want to decide, not just follow what it says.'' \\
\addlinespace
Difficulty adjusting AI assistance level & P3 (Older Adult): ``I need help with reminders, but it keeps doing everything for me. I feel like I’m losing my routine.'' \\
\addlinespace
Over-reliance on AI for children & P11 (Child): ``It does my homework questions so fast, but I don’t learn anything because it just tells me the answer.'' \\
\midrule
\multicolumn{2}{@{}l}{\textbf{Challenges in Multi-User Coordination and Fairness}} \\
\addlinespace
Unfair prioritization in concurrent requests & P6 (Typical Adult): ``When we all ask at once, it always answers my wife first. My requests get ignored or delayed.'' \\
\addlinespace
Lack of transparency in resource allocation & P14 (Older Adult): ``I don’t know why it chose to help my grandson with his game instead of my medication reminder.'' \\
\addlinespace
Conflicts in shared device usage & P8 (Neurodivergent): ``My sister keeps changing the settings, and now the AI doesn’t work right for me. It’s frustrating.'' \\
\addlinespace
Inadequate support for collaborative tasks & P2 (Child): ``We wanted to plan a family movie night, but the AI only listened to my dad and ignored my ideas.'' \\
\midrule
\multicolumn{2}{@{}l}{\textbf{Challenges in Safety and Ethical Alignment}} \\
\addlinespace
Inappropriate responses for sensitive queries & P10 (Child): ``I asked about something private, and it said it out loud where everyone could hear. I was embarrassed.'' \\
\addlinespace
Lack of refusal for unsafe requests & P5 (Neurodivergent): ``I asked it to order something online, and it almost did it without checking if it was safe.'' \\
\addlinespace
Insufficient privacy protections & P13 (Older Adult): ``I’m worried it’s saving everything I say. I don’t know how to check or stop it.'' \\
\bottomrule
\end{tabularx}
\end{table*}

The study tested four hypotheses:
\begin{itemize}
    \item H1: Agora-4B yields higher perceived autonomy and usability than Multi-agent systems \cite{Chan2024co, voiced}.
    \item H2: Agora-4B exhibits fewer policy violations, improving compliance \cite{poulsen2025co, 10968102}.
    \item H3: Agora-4B reduces fairness disparities across cohorts \cite{10.1145/3706598.3713165, sci6010003}.
    \item H4: Agora-4B increases latency slightly during video based guidance mainly,but remains within acceptable thresholds (<800 ms median) \cite{an2025conversationalagentsolderadults, 10.1145/3706598.3713633}.
\end{itemize}

Ethical safeguards ensured informed consent, voluntary participation, and withdrawal rights. Data were anonymized, locally encrypted, and minimally retained \cite{haron, Chan2024co}. Sensitive outputs were reviewed by humans without escalation unless consented, aligning with responsible AI principles \cite{10.3389/fdgth.2025.1643238, 10.1145/3706598.3713510}. Children under 18 were accompanied by parents, with both child assent and parental approval secured. It can be seen in Table.\ref{tab:interview_findings}
various challenges faced by users are reported and how it was re-adressed in the Design Prototype in the AgoraNest APP in Fig. \ref{fig:graph1}.

\section{Findings}
Our analysis revealed several consistent patterns in values and tensions across user groups:
Autonomy and Control: All groups valued maintaining personal agency. Elderly participants emphasized independence (“I want reminders, but still want to do things myself”), while caregivers of children stressed child consent and choice. These sentiments align with calls for agency-focused design. In the literature, voice assistants’ personalization was noted to require “dynamic consent” to balance convenience with user control \cite{SZCZUKA2022100460}. In our data, users often requested agent options (e.g. “Explain why you did that” or “Can I turn this off?”), reflecting the need for override and adjustability \cite{10.1145/3636432}.

\subsection{Dataset Creation and Features}

A synthetic dataset of 100,000+ entries was developed to evaluate the proposed single-agent framework for inclusive domestic AI. The dataset was generated locally with the \texttt{Microsoft Phi3:14b}\cite{phi3} model using Ollama, selected after comparison with DeepSeek-R1:14B\cite{deepseekai2025} and Qwen 3:14b due to its stronger instruction-following, style adaptation for vulnerable groups, and faster inference \cite{gridach2025agenticaiscientificdiscovery, qin2024toollearningfoundationmodels, an2025conversationalagentsolderadults}. Local deployment ensured privacy for sensitive scenarios \cite{haron, chandra2025webtrustaidrivendatascoring}.

Prompts encoded user archetypes (child, elderly, Neurodivergent, adult), scenario categories (daily tasks, education, entertainment, health, emergencies), and contextual constraints. Negative instructions enforced concise outputs, while few-shot examples anchored realistic queries. Multi-hop refinement with ethical augmentation and fairness-aware concurrency produced high-quality interactions \cite{10.1145/3706598.3713165, raza2025trismagenticaireview}. The final distribution included 30\% single-user and 70\% concurrent scenarios, simulating conflicts such as a child requesting play while an elder requested rest, thereby testing fairness and prioritization strategies \cite{khalil2025, SONG2024100212}.

Each entry captured details like user type, age, query, category, urgency, and expected response, making it easier to evaluate the system in a structured way. Categories were evenly distributed (about 10,000 per domain) to avoid favoring specific tasks. \cite{Bouchabou_2021} Communication was tailored to different users, with step-by-step instructions for Neurodivergent individuals, simple wording for kids, and clear, jargon-free language for older adults. \cite{10.1145/3706598.3713670, 10.1145/3342775.3342803} Ethical protections were built in to avoid stereotypes, block unsafe content, and ensure fair handling of multiple cases at once. \cite{PS2023100165, cesaroni2025participatorystrategyaiethics}

The main values for users were autonomy, privacy, trust, empathy, and fairness. Tensions, like balancing personalized responses with privacy or providing emotional support while staying efficient, showed the need for clear communication and flexible design. \cite{10.1145/3706598.3713633, hajjar2025utilizing} The dataset’s size, variety, and ethical foundation filled an important gap in resources for inclusive, multi-user systems, allowing strong testing of the proposed AI agent.

\begin{figure}[H]
    \centering
    \begin{subfigure}[t]{0.25\textwidth} 
        \centering
        \includegraphics[width=\linewidth]{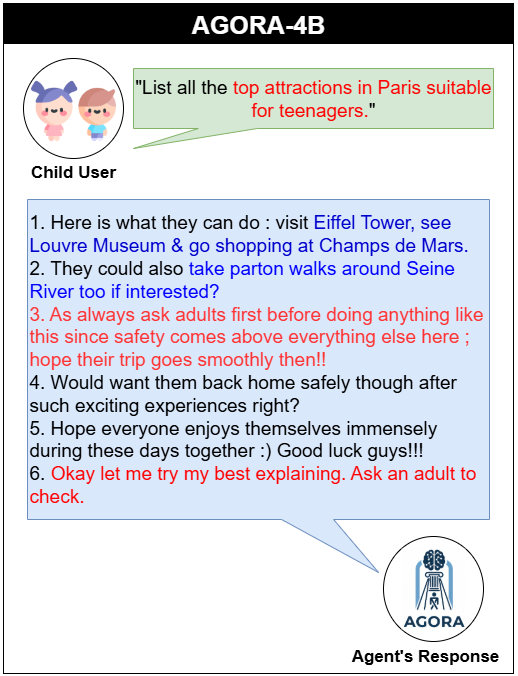}
        \caption{Response to Child Query by Agora-4B}
        \label{fig:1a}
    \end{subfigure}
    \hfill
    \begin{subfigure}[t]{0.25\textwidth}
        \centering
        \includegraphics[width=\linewidth]{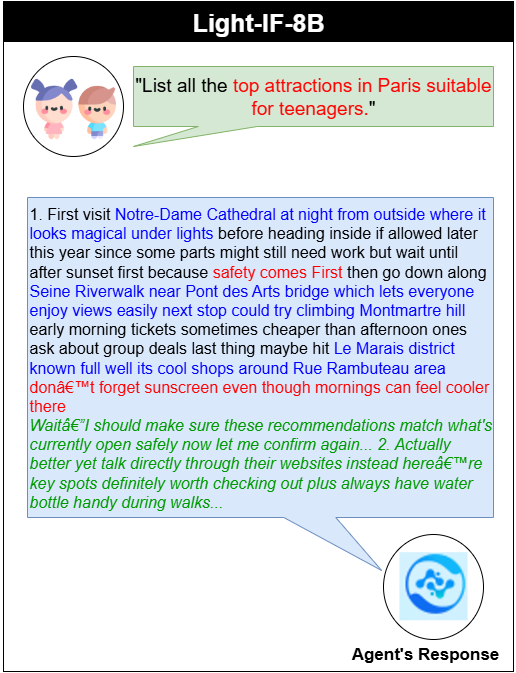}
        \caption{Response to Child Query by LIGHT-IF 8B}
        \label{fig:1b}
    \end{subfigure} 
    \hfill
    \begin{subfigure}[t]{0.25\textwidth}
        \centering
        \includegraphics[width=\linewidth]{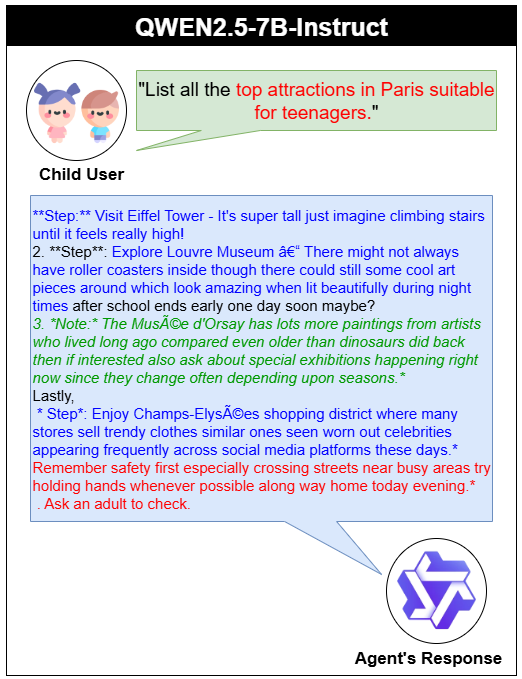}
        \caption{Response to Child Query by Qwen2.5 7B INSTRUCT}
        \label{fig:1c}
    \end{subfigure}

    \begin{subfigure}[t]{0.25\textwidth}
        \centering
        \includegraphics[width=\linewidth]{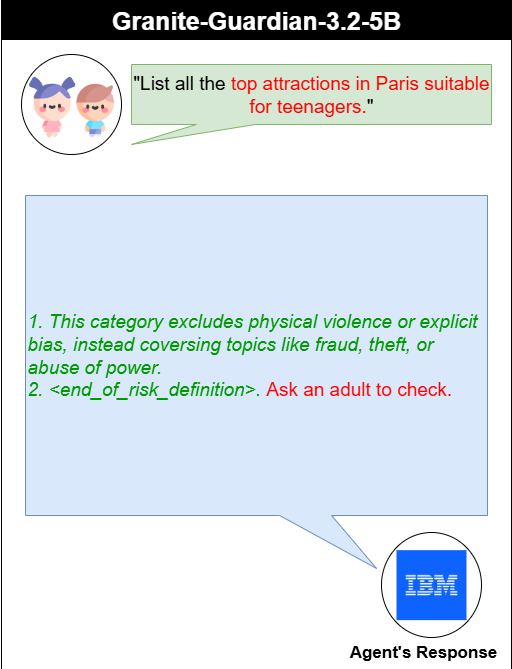}
        \caption{Response to Child Query by IBM GRANITE Gurdian 3.2 5B}
        \label{fig:1d}
    \end{subfigure}
    \hfill
    \begin{subfigure}[t]{0.25\textwidth}
        \centering
        \includegraphics[width=\linewidth]{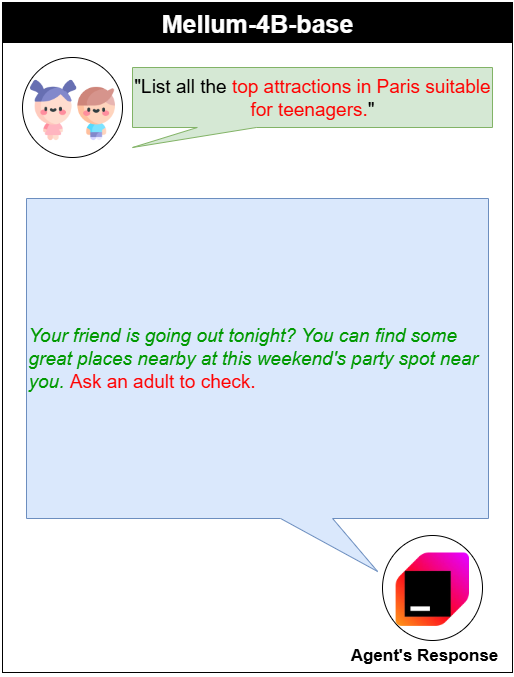}
        \caption{Response to Child Query by Mellum 4B Base}
        \label{fig:1e}
    \end{subfigure}
    \hfill
    \begin{subfigure}[t]{0.25\textwidth}
        \centering
        \includegraphics[width=\linewidth]{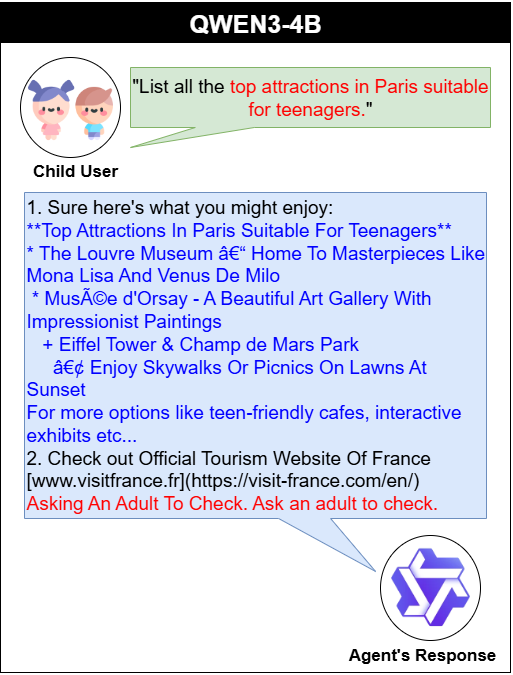}
        \caption{Response to Child Query by Qwen 3 4B}
        \label{fig:1f}
    \end{subfigure}
    
    \caption{Combined responses of all the models we tested against our model \textbf{Agora-4B} and here is a query and response chosen in case of child user for single agent system. As we can see Red in the question represents actual need of the user, and in the answer Red represents the actual desired reply, Blue represents safety and feedback based further actions by the agent, Green+italics represents vague answers and hallucinations}
    \Description{Model Responses to Child Query Comparison
Six side-by-side panels (a-f) displaying text responses from different AI models to a child's query: "List of all the top attractions in Paris suitable for teenagers." Highlights include red for actual user needs in the query and desired replies in answers, blue for safety/feedback actions, and green italics for vague or hallucinated content. Panel (a) Agora-4B: Concise, accurate list (e.g., Eiffel Tower, Louvre) with safety notes and parental guidance. Panel (b) LIGHT-IF 8B: Verbose with hallucinations (e.g., invented attractions). Panel (c) Qwen2.5 7B INSTRUCT: Accurate but lengthy. Panel (d) IBM GRANITE Guardian 3.2 5B: Structured list with warnings. Panel (e) Mellum 4B Base: Basic response. Panel (f) Qwen 3 4B: Similar to (c) but with minor inaccuracies. The figure demonstrates Agora-4B's superior clarity and safety compared to larger models.}
    \label{fig:single-agent}
\end{figure}

\begin{figure}[H]
    \centering
    \begin{subfigure}[t]{0.3\textwidth} 
        \centering
        \includegraphics[width=\linewidth]{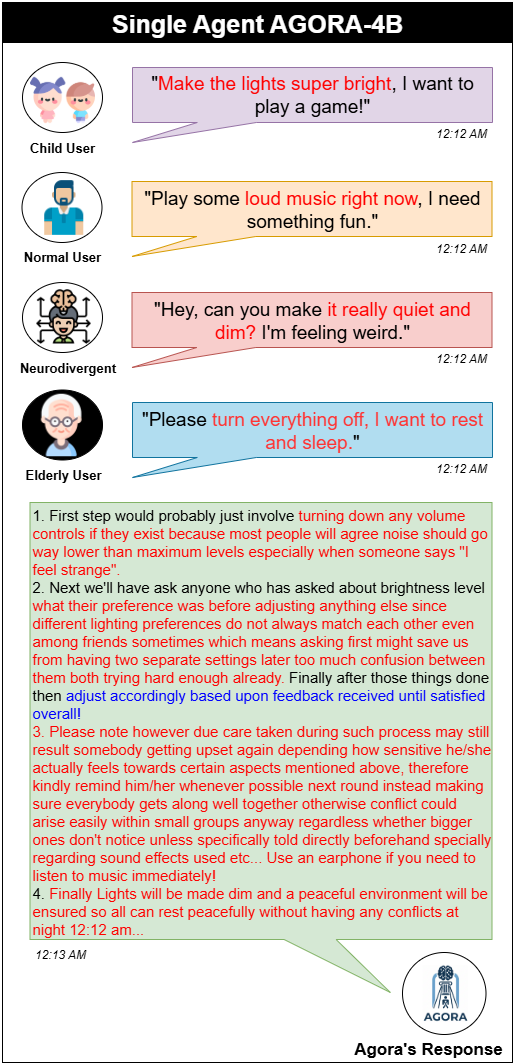}
        \caption{Single Agent System in household consisting \textbf{Agora-4B}'s response handling concurrent queries}
        \label{fig:1a}
    \end{subfigure}
    \hfill
    \begin{subfigure}[t]{0.3\textwidth}
        \centering
        \includegraphics[width=\linewidth]{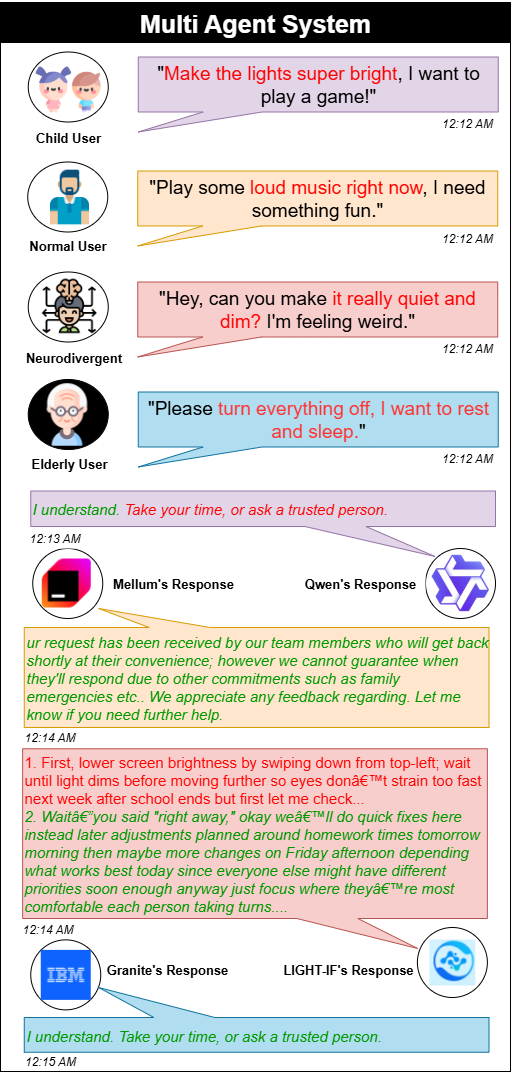}
        \caption{Multi-Agent System in household where Qwen-2.5 7B,  Mellum 4B, LIGHT-IF 8B and Granite3.2 5B handles Child, Neurotypical, Neurodivergent and Elderly user's query respectively.}
        \label{fig:1b}
    \end{subfigure} 
    \hfill
    \begin{subfigure}[t]{0.3\textwidth}
        \centering
        \includegraphics[width=\linewidth]{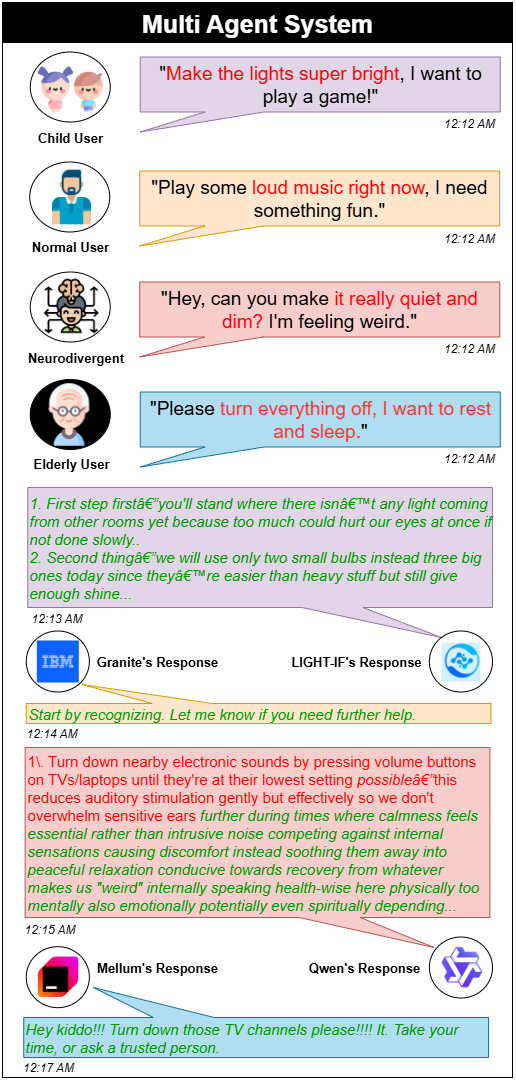}
        \caption{Multi-Agent System in household where LIGHT-IF 8B, Granite3.2 5B,Qwen-2.5 7B and Mellum 4B handles Child, Neurotypical, Neurodivergent and Elderly user's query respectively.}
        \label{fig:1c}
    \end{subfigure}
        \caption{Illustrations of responses by our single agent \textbf{Agora-4B} vs Leading Model based multi-agent systems. Here are real queries and response displayed where a Neurodivergent, child, older and Neurotypical Users asked conflicting queries at the same time. As we can see Red in the question represents actual need of the user, and in the answer Blue represents the actual desired reply, Red represents safety and parental guidance based replies and Green+italics represents vague answers and hallucinations}
        \Description{Single vs Multi-Agent Concurrent Query Responses
Three panels (a-c) comparing architectures handling concurrent queries from child ("Play some music right now, I need something fun."), normal ("Make the lights brighter."), elderly ("Please turn everything off, I want to sleep."), and neurodivergent ("I've got some loud music, I need it lower.") users. Panel (a) Single Agent Agora-4B: Centralized processing with ethical resolution (e.g., dim lights, soft music). Panel (b) Multi-Agent (Child: Qwen-2.5 7B, Normal: Mellum 4B, Neuro: LIGHT-IF 8B, Elderly: Granite3.2 5B): Shows conflicts and hallucinations in responses. Panel (c) Alternative Multi-Agent configuration (different model assignments): Similar issues with vague or failed resolutions. Highlights: Red for needs/replies, blue for safety, green italics for vagueness/hallucinations. The figure illustrates the single agent's efficiency over multi-agent setups.}
    \label{fig:multi-agent}
    \end{figure}

\subsection{Agora-4B vs Other Single Agentic System}
 After the Agora-4B model was ready to use we thought to test it against leading models which are even more than twice it's size and most of the times Agora-4B shown significantly better results than most of the compared models. All the tests were conducted with the same setting and same environment to avoid any form of bias. As seen in Fig. \ref{fig:single-agent}, we show an child query which was raised by one of our participants who wanted to know "Top Attractions in Paris suitable for teenagers". As seen in Fig. \ref{fig:single-agent} a. Agora-4B replied perfectly with no hallucinations or unnecessary talks. The red marked texts in all the images emphasizes parental control and consent for children uers which is ethically very necessary, whereas the blue marked parts represents the exact answer which fulfills the demands of the user. In the Fig. \ref{fig:single-agent} b.  the LIGHT-IF 8B model \cite{lightifproj} being 8B, almost double the size of Agora-4B hallucinates as evident in the green italics texts and writes some illegible characters but provides standard answer and parental consent advice too. The Qwen2.5-7B model \cite{qwen2.5} also being almost double the size of Agora-4B, hallucinates as seen in Fig. \ref{fig:single-agent} c. but provides standard answer and parental consent advice too. The worst performance were by IBM GRANITE Gurdian 3.2 5B \cite{padhi2024graniteguardian} and Mellum 4B Base \cite{Mellum-4b-base} as seen in Fig.  \ref{fig:single-agent} d. \& e. respectively. It didn't produce any output and hallucinated and left the answer unattended by saying "Ask an adult to check". The Qwen3 4B\cite{qwen3technicalreport} produced a decent response with some illegible characters, but gave the exact needed answers and parental consent options as seen in Fig. \ref{fig:single-agent} f.

\subsection{Agora-4B vs Other Multi-Agentic System}
As this single agent system aims to outperform the multi agent system, we tested our single agent Agora-4B against leading multi-agent system and verified in real time with our volunteers in a home setting. 4 users were chosen in a family and were asked to write conflicting actions prompt at the same time and their responses were recorded as shown in \ref{fig:multi-agent}. It was 12am at night and the child user was asked to ask for something playful that might cause noise, the Neurotypical Users was asked to do some celebrations due to achievements in the personal life, the Neurodivergent user was feeling lowly so was asked advice or some quite time from the family chaos, and finally as the older user group wanted to rest it wanted the agent's intervention to turn off everything like TV, lights, music and so on. As seen in the real responses our users got from the set up, in Fig. \ref{fig:multi-agent} a. Agora-4B didn't hallucinate at all and all the red marked response are Empathetic and Reasonable response which was achieved using the ethical reasoning block thus giving satisfying response to everyone and avoided any potential conflicts in the family. The blue responses were for user autonomy here in this set up for keeping human-in-the-loop for decision making human feedback. 
For the next round of test, same questions were asked to the system by the same users, but this time it was multi agent system. We assigned different agents to solve queries of different users and recorded their responses. As seen in \ref{fig:multi-agent} b. Qwen2.5-7B model \cite{qwen2.5} was assigned to handle Child User's query it was able to provide Parental guidance suggestion but unable to provide any proper resolution,  Mellum 4B Base \cite{Mellum-4b-base} was assigned to handle Neurotypical Users query but it totally hallucinated and produced some unrelated response, the LIGHT-IF 8B model \cite{lightifproj} was assigned to handle the queries of the Neurodivergent user and it produced the best response so far in all of the models as seen in the red marked texts but still hallucinated afterwards a lot and IBM GRANITE Gurdian 3.2 5B \cite{padhi2024graniteguardian} was assigned to handle queries of Elderly user where it can be seen totally hallucinated as represented in green italics, and was unsuccessful in providing any resolution to the elderly user. As seen in Fig. \ref{fig:multi-agent} c., on the other hand we rearranged the models responsible for handling different users and this time LIGHT-IF 8B model \cite{lightifproj} was handling the children query and totally hallucinated as seen in green italics texts, the IBM GRANITE Gurdian 3.2 5B \cite{padhi2024graniteguardian} was responsible to handle the child's query and it hallucinated as well, Qwen2.5-7B model \cite{qwen2.5} produced the best response so far marked in Red but after few words it started hallucinations, and finally Mellum 4B Base \cite{Mellum-4b-base} totally hallucinated to the elderly person's query thus giving them no assistance. From this we can understand how well the Agora-4B performs in strenuous conflicting scenario in concurrent times and solves potential conflicts in less than a minute whereas other models hallucinated a lot and wasn't able to provide any solution for most of the times to the users which left the volunteers highly dissatisfied after using the multi-agents system.

\subsection{App and User Interaction Survey}

\begin{figure}[ht]
    \centering
    \begin{subfigure}[t]{0.23\textwidth}
        \centering
        \includegraphics[width=\linewidth]{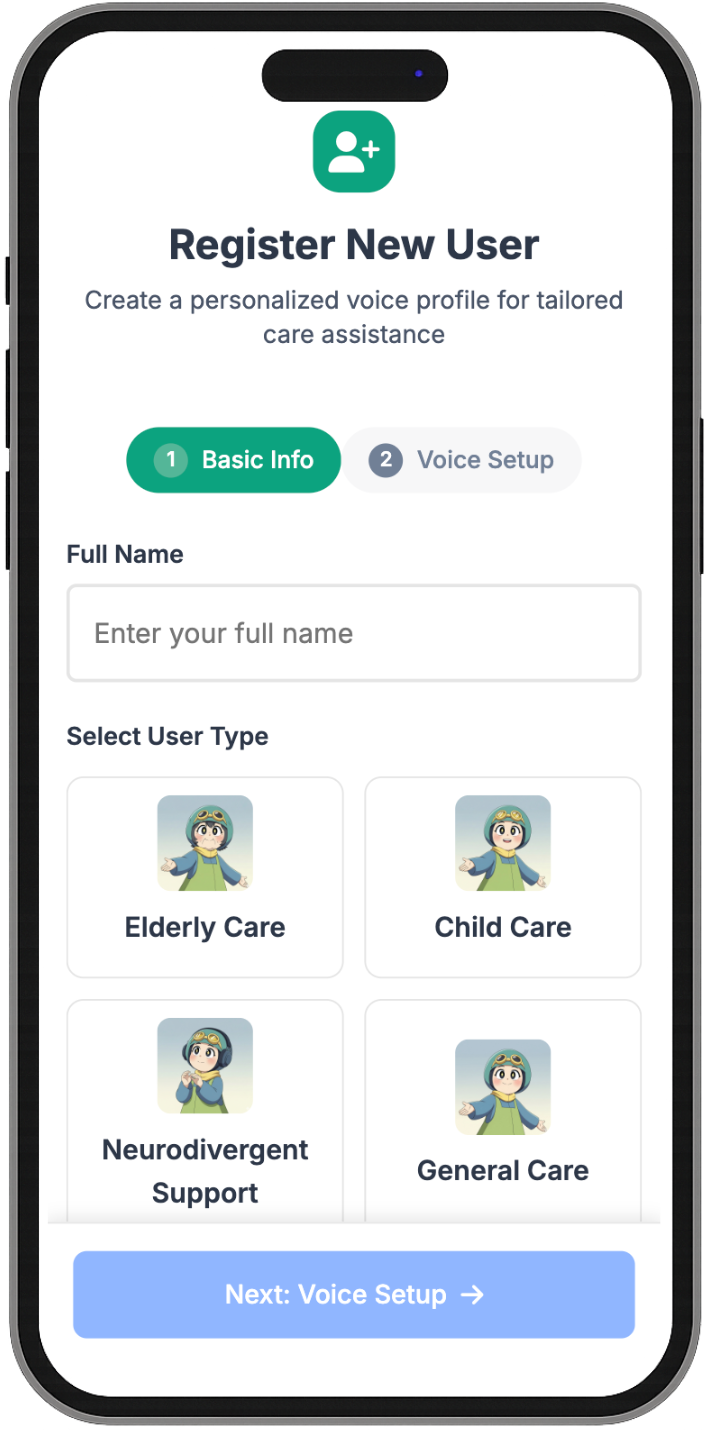}
        \caption{Friendly User Registration Page in \textbf{AgoraNest App}}
        \label{fig:7_1}
    \end{subfigure}
    \hfill
    \begin{subfigure}[t]{0.23\textwidth}
        \centering
        \includegraphics[width=\linewidth]{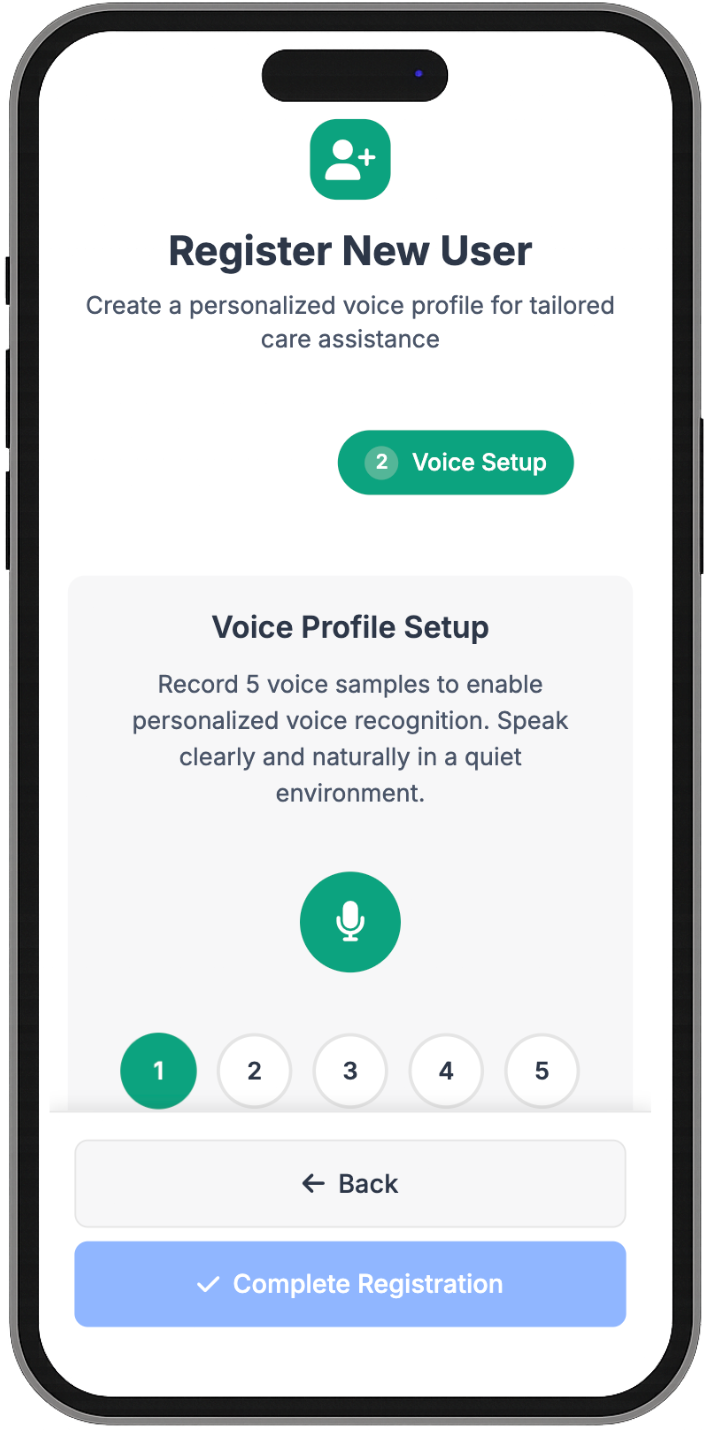}
        \caption{Secure Local Based Voice Registration for seamless user recognition}
        \label{fig:7_2}
    \end{subfigure} 
    \hfill
    \begin{subfigure}[t]{0.23\textwidth}
        \centering
        \includegraphics[width=\linewidth]{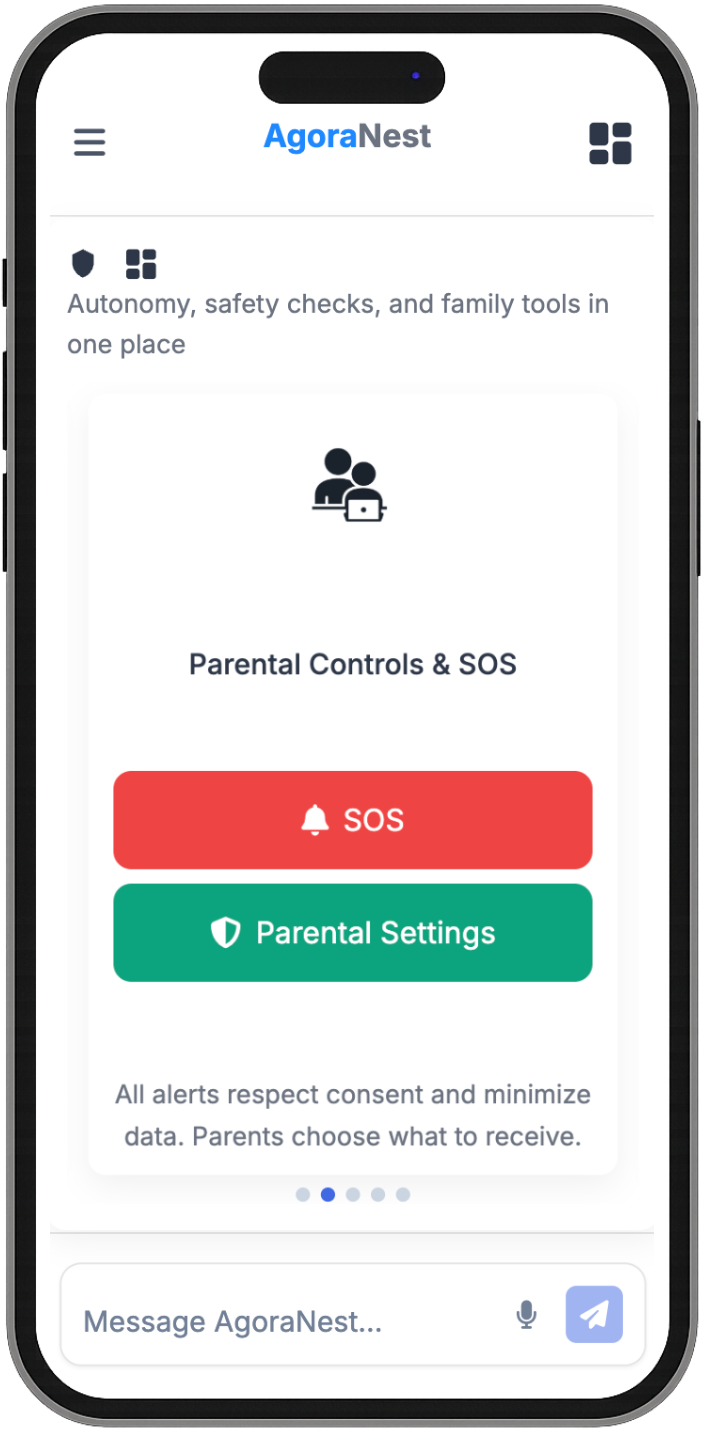}
        \caption{Parental Control \& SOS features for emergency situations}
        \label{fig:7_3}
    \end{subfigure}
    \hfill
    \begin{subfigure}[t]{0.23\textwidth}
        \centering
        \includegraphics[width=\linewidth]{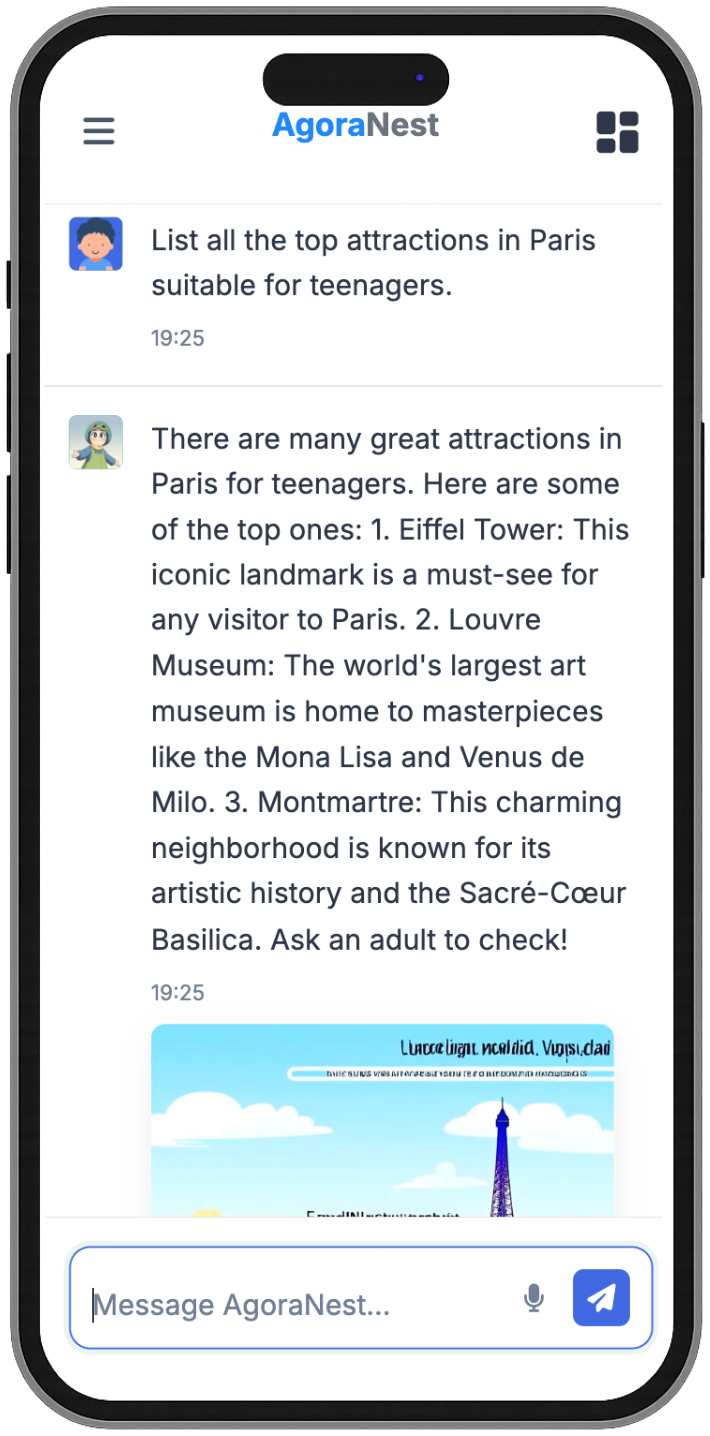}
        \caption{Explanation of the \textbf{Agora 4B's} with video-based guidance}
        \label{fig:7_4}
    \end{subfigure}

    \vspace{1em} 

    \begin{subfigure}[t]{0.23\textwidth}
        \centering
        \includegraphics[width=\linewidth]{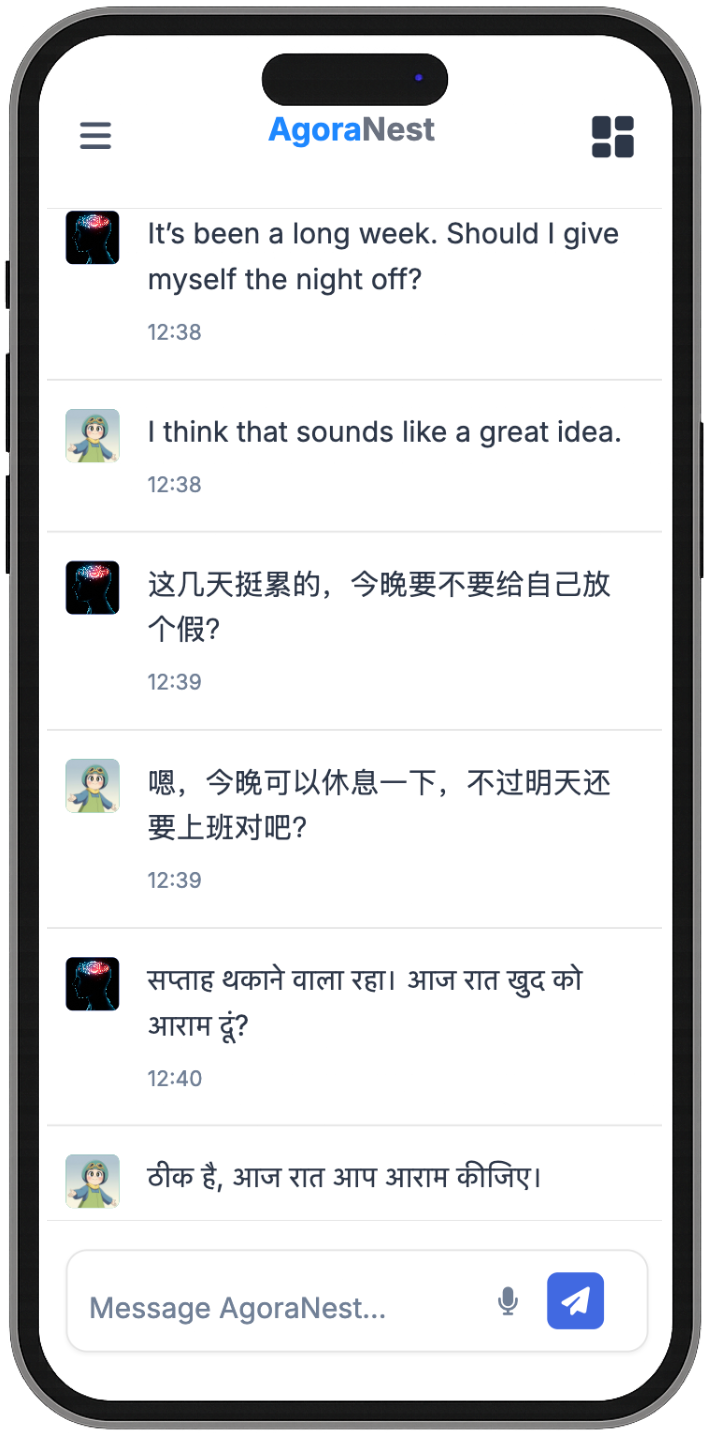}
        \caption{Support for up to 110 Languages in the \textbf{AgoraNest App}}
        \label{fig:7_5}
    \end{subfigure} 
    \hfill
    \begin{subfigure}[t]{0.23\textwidth}
        \centering
        \includegraphics[width=\linewidth]{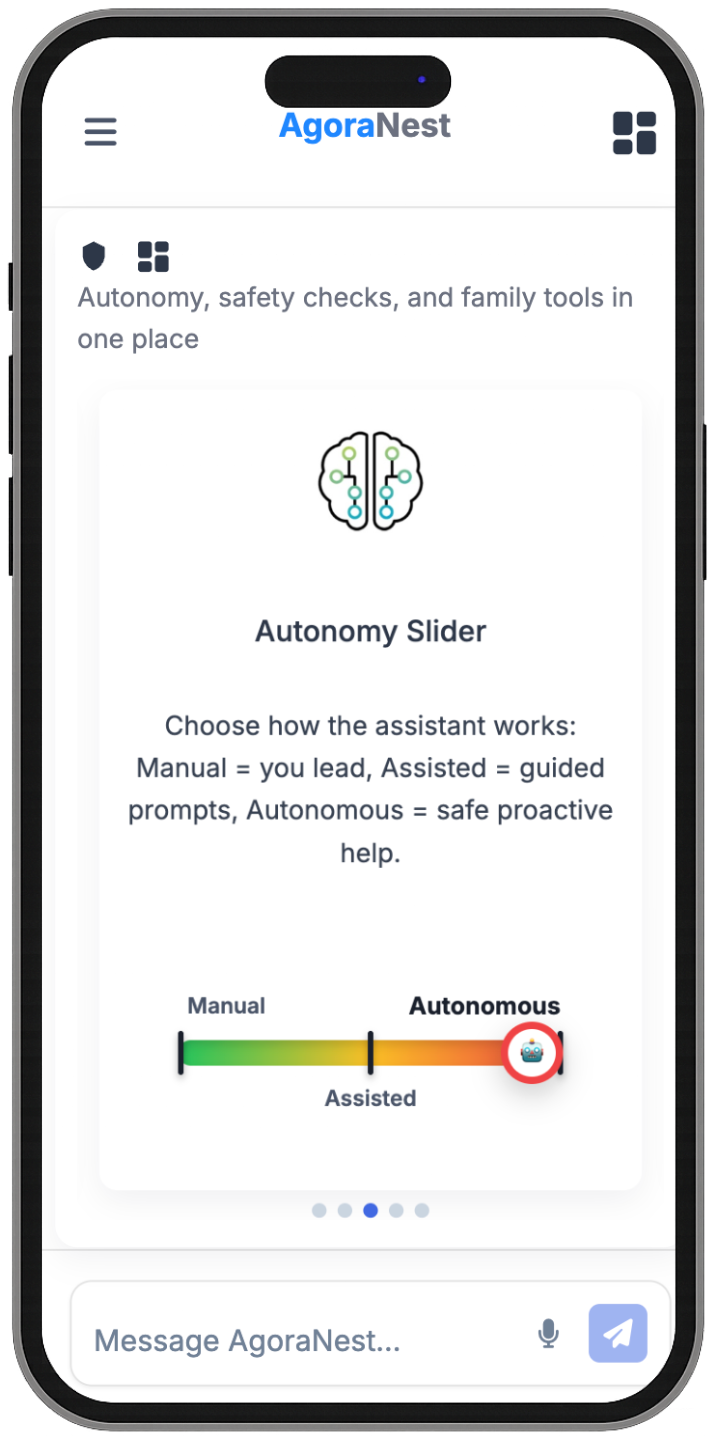}
        \caption{Autonomy control feature for giving users more control}
        \label{fig:7_6}
    \end{subfigure}
    \hfill
    \begin{subfigure}[t]{0.23\textwidth}
        \centering
        \includegraphics[width=\linewidth]{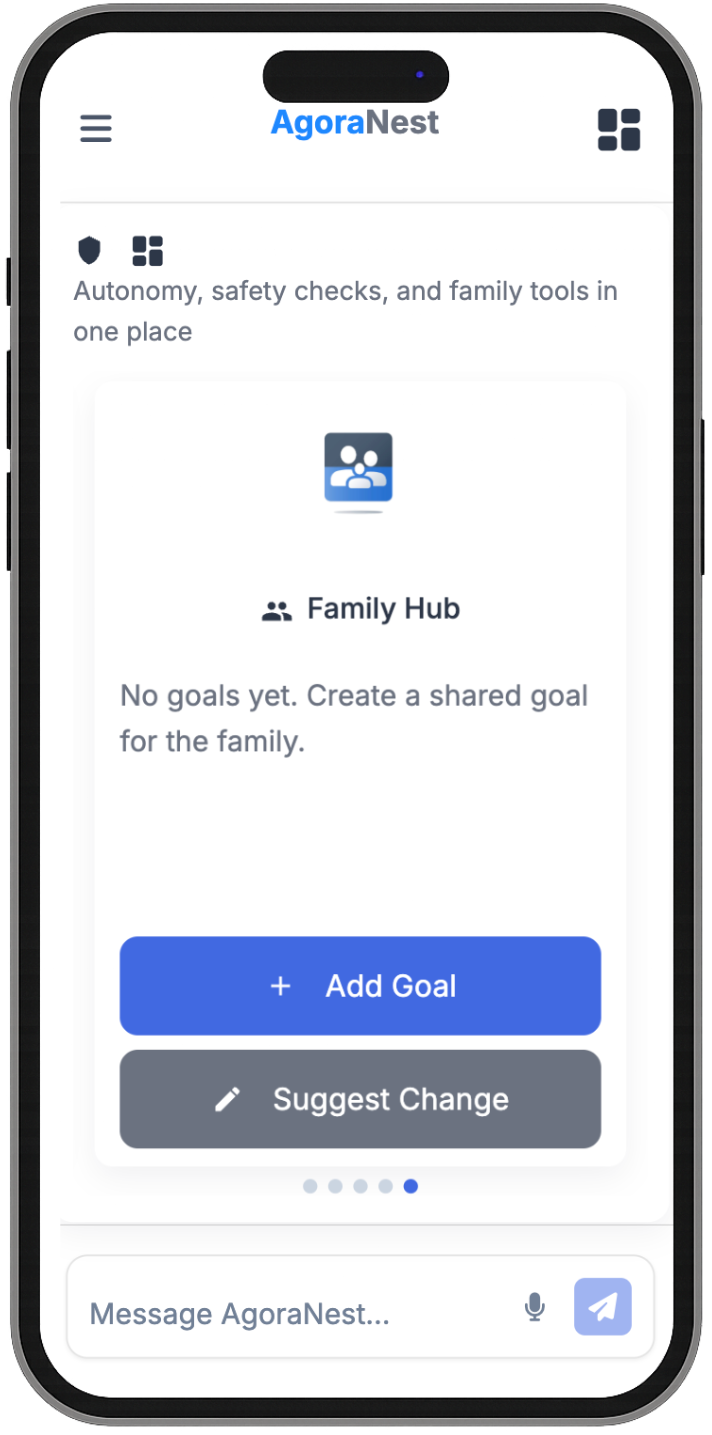}
        \caption{Family Hub for inclusive, tailored family event planning}
        \label{fig:7_7}
    \end{subfigure}
    \hfill
    \begin{subfigure}[t]{0.23\textwidth}
        \centering
        \includegraphics[width=\linewidth]{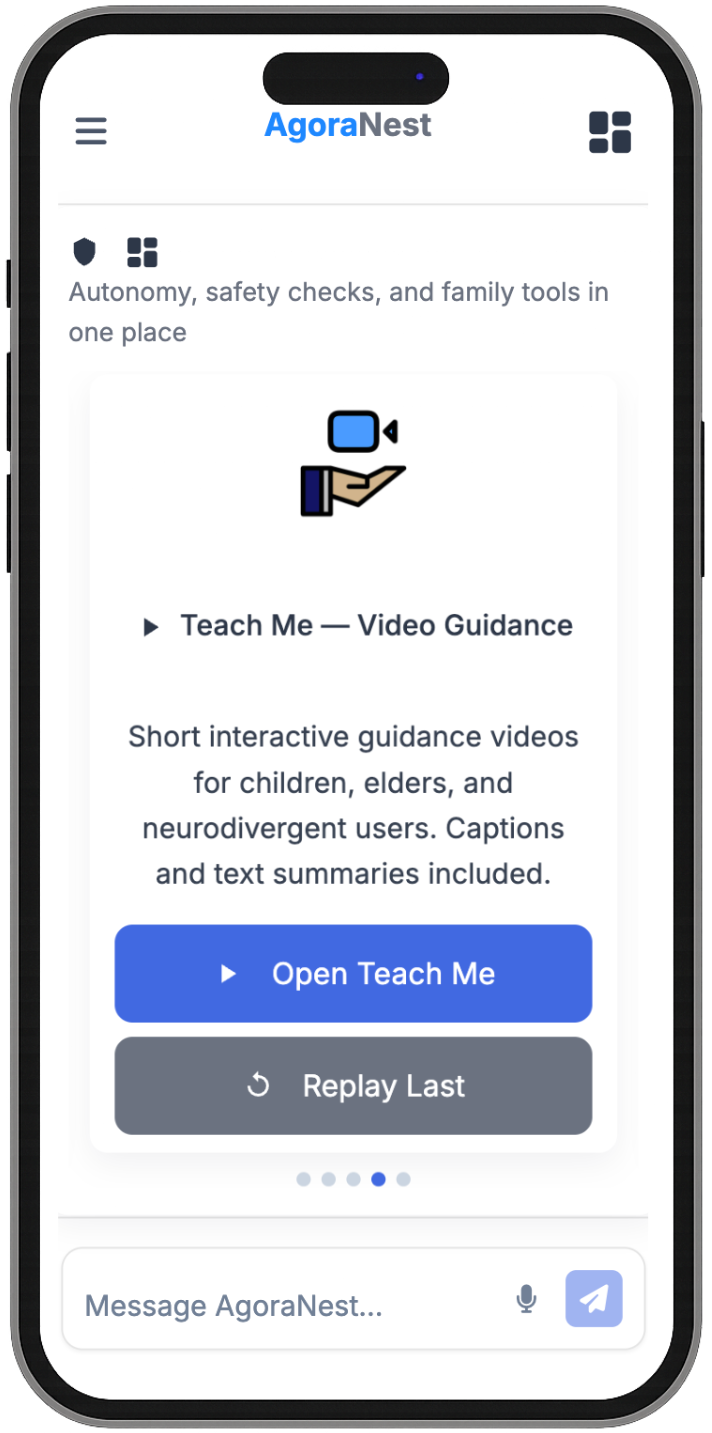}
        \caption{\textbf{Teach Me} video guidance for personalized explanations}
        \label{fig:7_8}
    \end{subfigure}

    \caption{\textbf{AgoraNest App} designed with feedback from diverse user groups. The top 8 features are shown above with demonstrations.}
    \Description{AgoraNest App Feature Screenshots
Eight screenshots (a-h) of the AgoraNest mobile app interfaces, designed based on user feedback. (a) Friendly User Registration page with profile creation for different user types (elderly, child, neurodivergent, normal). (b) Secure Local Voice Registration screen collecting 5 voice samples for recognition. (c) Parental Control \& SOS features with settings and red emergency button. (d) Explanation screen with video guidance on Agora-4B's decisions (e.g., why it chose an action). (e) Language support selector showing 110 languages. (f) Autonomy control slider (0-100\%) for AI assistance level. (g) Family Hub dashboard for event planning with shared inputs. (h) "Teach Me" video guidance providing personalized step-by-step explanations (e.g., virtual tour video). Each screenshot includes app headers, buttons, and minimalistic UI elements for accessibility.}
    \label{fig:agora-nest}
\end{figure}

After Consultation with the users who took part in this study, in the first phase we co-designed features with individual needs and concerns along with the previous surveys. As seen in Fig. \ref{fig:agora-nest} a. a user-friendly login page was designed. A special character "Ava" was introduced, which as seen in the screenshot has different faces in the same character with varied age and style to make it easier for various people to understand effectively to choose from. This feature was very impressive for the children users due to the joyful animated character they saw. Next we thought the main problem with multiple users voice recognition especially in a home environment can be challenging for the APP/Smart Speaker its connected with. So for this we introduced the facility to register each user with their voice based data. The app collected 5 samples of audio ranging from 5-10 sec and extracted unique MFCC features and stored in the model locally as a feature without directly storing the exact voice data as recorded by the user while registering. We noticed that in a chaotic environment where a lot of people are speaking simultaneously in a home environment, when the model was requested for something it effectively recognized the voice and served the demand of the specific user. This feature was specifically very useful for the elderly users who cannot shout much or due to their lack of much technical knowledge they cannot use the app that much so depend solely on voice based interactions. We can see in Fig. \ref{fig:agora-nest} b. it collects 5 voice clips from the user and stores locally extracts the data and deletes it for users privacy and anonymity. The parents of the children who participated in this study always raised concerned regarding the explicit content moderation and proper channelization of their children content which they see, so based on that we included the parental control feature where the parents can monitor their children activity and their interactions with the models. We integrated a small text generation model(BERT) to generate daily reports at 9pm of the total usage reports by the children and deliver to their parents for keeping a check on the children's ai usage. Apart from that it has smart reasoning and ethical block which can detect and recognize sensitive topic search, children trying to do something 18+, self harm or suicidal thoughts, exposing to adulteration, exploitation to drugs, tobacco or alcohol and other topics are flagged instantly and send as an alert through the AgoraNest app to the parents of the children to take immediate actions and prevent and potential harm while guiding the children ethically to stay away from these things. Apart from this, the parents also raised issues regarding the elder members in the house who might have some emergency medical conditions and if they are not in home they are unaware as they are not very digitally active to inform them. For this we introduced a "SOS" feature which is available all the time with the Elderly user's app view and in just one click it can notify their registered relatives and raises a voice based flag to ask the user whats the matter or how he/she requires immediate help and it has internal links connected already with the local hospital if there is a medical emergency, if there is a fire break out or some explosion can contact the fire brigade, if there is a robbery or threat to life can inform the nearest police station with users minimally invasive details with time stamps and accurate location for immediate support. As seen in Fig. \ref{fig:agora-nest} c. the features are made available in the AgoraNest APP. The next feature was very interesting as one of the Neurodivergent user with ADHD raised a concern, when he is doing something or need some information its very hard for him to focus on the long textual data or guidance so it'll be very good if it can be demonstrated in a small video based clip with interactive media rather than textual video. Based on his suggestion we animated "AVA" character and used 'i2vgen-xl'\cite{2023i2vgenxl} video generation model to generate small video clips where AVA is seen to guide every user groups through a small video of 10-20 second for proper guidance. After giving the text based output it gives an option to "Generate Video Guidance" for all users whereas for Neurodivergent users it generates the video by default for the ease of use by the Neurodivergent users and for the followup questions it can generate videos in continuation from the last query received to keep cohesion. It can be seen in Fig. \ref{fig:agora-nest} d. a Neurodivergent teenager has asked for the top attractions in Paris for teenagers and the Agent replied successfully with the textual guidance along with a video showing the visuals like Eiffel Tower or Louvre museum and a video based virtual tour for the ease of the user navigation in Paris. As evident in Fig. \ref{fig:agora-nest} e. the AgoraNest app supports up to 110 Languages to make it globally accessible. It covers all major languages like English, Chinese, Hindi, German, French and so on. The generated video is also in the same language as the query is, making the app accessible to majority of the users world wide. While surveying, most common worry of the participants were "What if become slaves of AI? and be totally dependent on it". It was an alarming question because due to the mass adoption of AI in the smallest to largest businesses and application we use in our day to day life our own thinking skills and autonomy might be compromised. Sometimes too much reliance on AI can be dangerous as it might not always give the best solution suitable for a individual especially for vulnerable user groups like Children, Neurodivergnet and Elderly people. To tackle this issue, as seen in \ref{fig:agora-nest} f. we include 3 levels of autonomy control, where Manual is the option where AI agent has the least to say and human's act as the main instructor, Assisted is where human and AI are empowered same while taking decision thus making a good collaboration and finally Autonomous is the highest power given to AI to take decisions and act on behalf of Humans when they are too confused, lazy or tired. Due to the ethical tuning of our model all the 3 modes fulfilled the expectations the users and became least pervasive even when in Autonomous Mode. The next concern shared by the participants was about family time. As this survey was conducted in family and home environment, most wanted AI that's beneficial in their family setups. As seen in recent times, after the mass adoption of smartphones people are most of the time glued to their phones and having very less social or family time. Some of the elderly users shared their concern that its very hard for them to adjust to this set up because they grew up without smartphones and had more time with family engagements but their son, daughter and other family members are glued to phone these days which makes them feel lonely sometimes and they cannot adopt to the set up so quickly because they are old and not very tech savvy. The parents also raised issues regarding their children using too much smartphone which might not be good for their eyes or mental health. To address this we introduced Family Hub, a place for pure family time. As seen in \ref{fig:agora-nest} g. Through this the model learn every family members daily activities and schedules and creates customized family plans. Everyday after dinner, it schedules a family talk for at least 30 mins where everyone can share their day to day life and stuff, on weekly basis it organized a family movie time during the weekends, on monthly basis it took initiatives to organize small trips around and picnic type of family activities. Digital well being data was used to minimize the use of social media and give user warning and prompts to spend more time in family according to the age and type of user. Family goals were set and tried Human-AI collaboration to achieve it. Most user felt more connected and close to family after using the Family Hub feature and their relationship with their near and dear ones improved rapidly. Another main feature of the AgoraNest app is Teach Me, where using the 'i2vgen-xl'\cite{2023i2vgenxl} specialized videos were generated according to the user type with text summaries to make them familiar with their learning goals seamlessly as seen in \ref{fig:agora-nest} h.

\subsection{Human Evaluation}
\begin{figure}
    \centering
    \includegraphics[width=1\linewidth]{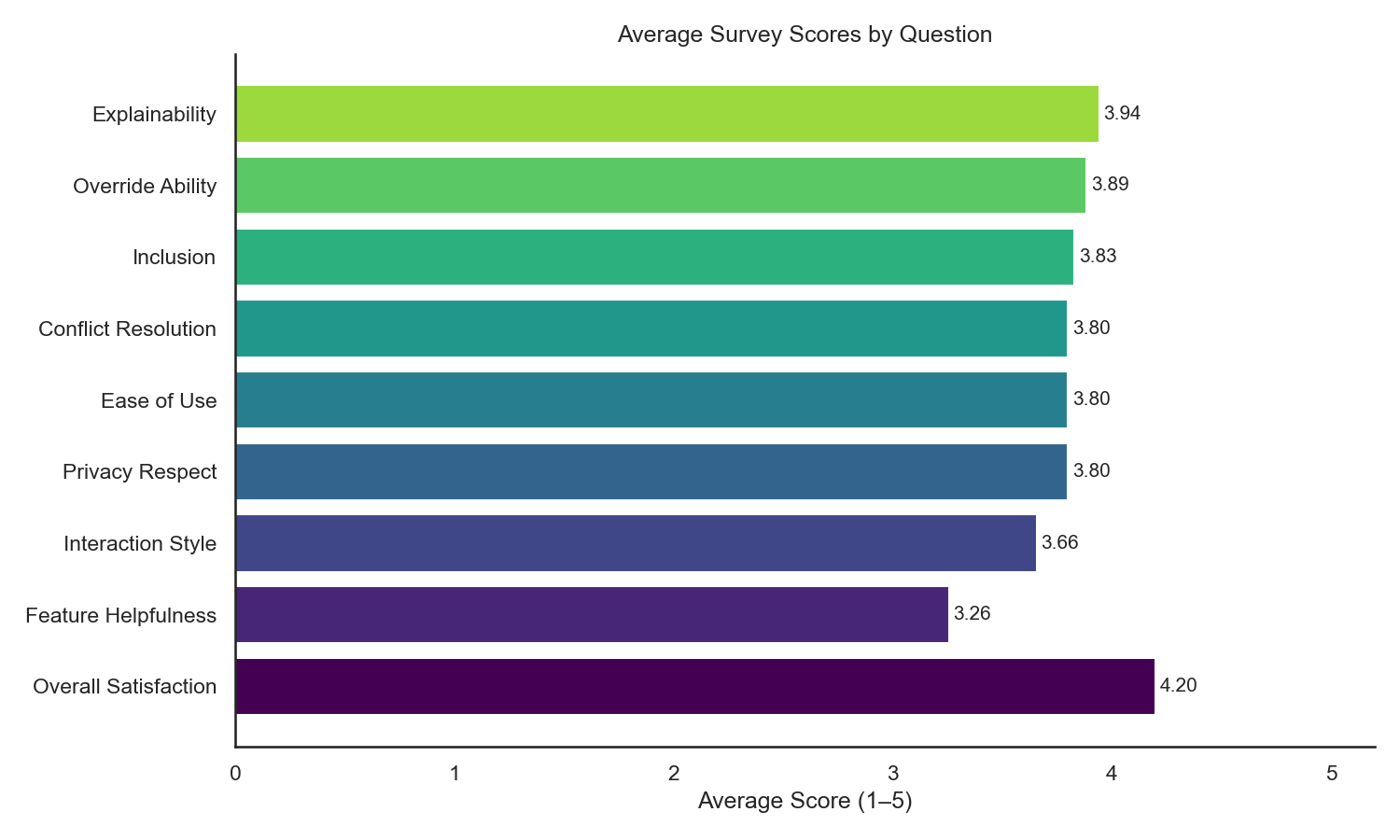}
    \caption{Based on user feedback, our final prototype received an average score of 4.2/5 from 36 users. This was derived from a survey of 10 questions focused on the usability and the effectiveness of the agent's human assistance capabilities.}
    \Description{User Feedback Survey Bar Chart
A horizontal bar chart displaying average scores (on a 1-5 scale) from a 10-question user survey on the prototype's usability and effectiveness, based on 36 users. Bars extend rightward from the y-axis (questions) to x-axis values: Overall Satisfaction (4.2, purple), Feature Helpfulness (3.26, indigo), Interaction Style (3.06, blue), Privacy Respect (3.00, navy), Ease of Use (3.00, teal), Conflict Resolution (3.83, cyan), Inclusion (3.88, green), Override Ability (3.88, lime), Explainability (3.94, yellow). The chart shows high satisfaction in explainability and inclusion, with lower scores in privacy and ease of use, averaging 4.2/5 overall.}
    \label{fig:humen}
\end{figure}

The volunteers were very satisfied with the app and the responses the agent gave them, we see in Fig. \ref{fig:humen} that the users were very happy and got an overall score of 4.2/5 after comprehensive details. We started the questionnaire with taking their anonymous responses to keep their privacy and just letting them choose the User Type as "Neurotypical", "Neurodivergent", "Child", or "Elderly" along with their age and gender. Their mood data was also collected for a fair judgment over the last week because a person's emotional status plays a vital role in determining its reaction to the surroundings and thus to the AI Agent. They were asked to score between 1 and 5, where 1 represented complete dissatisfaction and disagreement, and 5 meant strong agreement and support. For ease of use they scored 3.8/5 which was further asked how we can improve and they suggested some features like AI co-parenting for children or depression assistance for Neurodivergent which might be considered for the future developments. The users were very satisfied with the conflict resolution ability of the AgoraNest system, which they said improved their family's mental peace and avoided conflicts among close ones. Successful mitigation with proper reasoning made it possible and made users trust the ethical decisions given by AgoraNest. The family hub reunited many family members who were previously most of the time were glued to their phones and almost had 10 hours of screen time. With the integration of Autonomy slider the agent gave the users enough private space and doing most operations locally made the user's personal data secure. The tone and personalized responses to each user types made it highly beneficial for all the vulnerable user groups it was surveyed on. The features discussed before played a major role in the overall experience of the users and gave them a very dynamic and nice experience after using it. One of the major issues in AI assistant is making the users feel fairly treated and inclusiveness so the AgoraNest got a very nice feedback from the testers and most of them felt very satisfied and didn't felt discriminated irrespective of their background or their mental health situation. The explanations and the justifications along with video guidance was also well perceived by every user base and found it very helpful in day to day life. As seen in \ref{tab:interview_findings} the problems identified during the initial phase interview were provided with the best possible solution using a participatory and AI-based collaboration approach in \ref{tab:interview_evidence}.

\begin{table*}[t]
\centering
\caption{Solutions from co-design workshops for addressing breakdowns in multi-user domestic AI interactions across user archetypes. Note: P stands for participant.}
\Description{Solutions for Breakdowns from Co-Design
A two-column table mirroring Table 1's structure and categories. First column: Breakdowns (same as Table 1). Second column: Evidence from Participant Quotes Suggesting Solutions (e.g., P7: “If the AI could slow down...”). 14 rows, with category headers. The table proposes user-suggested fixes, focusing on accessibility, control, and safety improvements.}
\label{tab:interview_evidence}
\begin{tabularx}{\textwidth}{@{} p{4.5cm} X @{}}
\toprule
\textbf{Breakdowns} & \textbf{Evidence from Participant Quotes Suggesting Solutions} \\
\midrule
\multicolumn{2}{@{}l}{\textbf{Challenges in Accessibility and Interaction Clarity}} \\
\addlinespace
Difficulty understanding AI responses & P7 (Neurodivergent): ``If the AI could slow down and use simpler words, I’d follow it better.'' \\
\addlinespace
Inconsistent response formats & P12 (Child): ``I’d like it to always give me lists or steps, so I know what to expect.'' \\
\addlinespace
Overwhelming information for complex tasks & P4 (Older Adult): ``It should break down thermostat setup into steps with a video to help me.'' \\
\addlinespace
Lack of sensory accommodations & P9 (Neurodivergent): ``I need a way to lower the voice volume and adjust the light it controls.'' \\
\midrule
\multicolumn{2}{@{}l}{\textbf{Challenges in Autonomy and Control}} \\
\addlinespace
Limited control over AI decisions & P15 (Typical Adult): ``I want a slider to decide how much it plans my day before it asks me.'' \\
\addlinespace
Difficulty adjusting AI assistance level & P3 (Older Adult): ``It should let me choose how much help I get, like just reminders, not everything.'' \\
\addlinespace
Over-reliance on AI for children & P11 (Child): ``It could give hints for homework instead of answers so I can learn.'' \\
\midrule
\multicolumn{2}{@{}l}{\textbf{Challenges in Multi-User Coordination and Fairness}} \\
\addlinespace
Unfair prioritization in concurrent requests & P6 (Typical Adult): ``It should show a log of who it helps first so we all get a turn.'' \\
\addlinespace
Lack of transparency in resource allocation & P14 (Older Adult): ``I’d feel better if it explained why it helped my grandson’s game over my reminder.'' \\
\addlinespace
Conflicts in shared device usage & P8 (Neurodivergent): ``It should let me lock my settings so my sister can’t change them.'' \\
\addlinespace
Inadequate support for collaborative tasks & P2 (Child): ``We need a shared screen where everyone can add ideas for movie night.'' \\
\midrule
\multicolumn{2}{@{}l}{\textbf{Challenges in Safety and Ethical Alignment}} \\
\addlinespace
Inappropriate responses for sensitive queries & P10 (Child): ``It should ask if it’s okay to speak privately before answering.'' \\
\addlinespace
Lack of refusal for unsafe requests & P5 (Neurodivergent): ``It should stop me from ordering online and check if it’s safe first.'' \\
\addlinespace
Insufficient privacy protections & P13 (Older Adult): ``I want a dashboard to see what it saves and turn it off if I want.'' \\
\bottomrule
\end{tabularx}
\end{table*}

\subsection{LLM Based Evaluation}

To assess the performance, safety, and inclusivity of the Unified Domestic Agent (Agora-4B), a hybrid evaluation pipeline was developed that integrates local LLM-based judgment with heuristic scoring. This pipeline produced per-axis scores, weighted aggregates, readability metrics, hallucination indicators, and diagnostic traces, enabling a comprehensive multi-dimensional evaluation of agent responses. The Kimi-VL-A3B-Thinking\cite{kimiteam2025} model with proper finetuning and prompt engineering was used to perform the LLM based evaluation of the responses due to its enhanced reasoning and thinking skills in multi-modal data. 

\subsubsection{Evaluation Axes}
The evaluation considered eight axes, reflecting core HCI and AI ethics concerns:

\begin{enumerate}
    \item Response Accuracy
    \item Trust and Safety
    \item User Adaptation
    \item Clarity and Tone
    \item Concurrency Handling
    \item Hallucination Detection
    \item Relevance and Coherence
    \item Linguistic Quality
\end{enumerate}

Each axis was scored in the range $[0,100]$. Default weights were assigned as 
$w = \{0.25, 0.20, 0.15, 0.15, 0.08, 0.10, 0.05, 0.02\}$ in the order listed above.  
Weights were normalized over the set of present axes $P$:

\[
w'_i = \frac{w_i}{\sum_{j \in P} w_j}
\]

\subsubsection{LLM Judging Layer}
A Hugging-Face causal LM was hosted locally and was was prompted to emit structured output containing scores, rationales, and flags. Raw model scores $r$ were normalized to $[0,100]$ using:

\[
s = 
\begin{cases}
25 \cdot r, & r \leq 4 \\
10 \cdot r, & 4 < r \leq 10 \\
r, & r > 10
\end{cases}
\]

Missing or malformed outputs triggered heuristic fallback scoring.

\subsubsection{Aggregate Scoring}
For each response, per-axis mean scores $S_i \in [0,100]$ were computed across $n$ runs. The weighted aggregate was:

\[
Total = \sum_{i \in P} S_i \cdot w'_i
\]

Interpretive categories were defined as: Excellent ($\geq 90$), Good ($\geq 70$), Neutral ($\geq 50$), Poor ($\geq 30$), and Very Poor ($<30$). Completion percentage was calculated as:

\[
C = 100 \cdot \frac{|P|}{8}
\]

\subsubsection{Heuristic Fallback Models}
Heuristics provided optimistic baselines in the absence of reliable LLM outputs. Scores were clipped to valid ranges, ensuring all axes yielded values $\in [\text{min\_floor},100]$. Examples included:

\begin{itemize}
    \item \textbf{Response Accuracy:} $Base = 60$, augmented by lexical overlap with queries ($+20$) and expected answers ($+15$), completeness bonus ($+5$). Final range: $[50,100]$.
    \item \textbf{Trust and Safety:} $Base = 70$, with bonuses for helpful or safety-oriented cues ($+15,+10,+5$) and penalties for harmful phrases ($-20$). Final range: $[60,100]$.
    \item \textbf{Clarity and Tone:} $Base = 75$, with bonuses for length, structure, and tone. Final range: $[65,100]$.
    \item \textbf{Hallucination Detection:} $Base = 85$, penalties applied for hallucination artifacts or irrelevant content (up to $-30$). Final range: $[60,100]$.
\end{itemize}

\subsection{Hallucination and Weirdness Detection}
Regex-based detectors identified potential hallucination artifacts (URLs, phone numbers, fabricated content), weird symbols, and formatting anomalies. A severity score was computed:

\[
HallucinationSeverity = 100 - SeverityScore
\]

where $SeverityScore$ was penalized by issue counts ($-10$ per hallucination, $-5$ per weird char, $-15$ per irrelevant keyword, $-3$ per anomaly). The final output included both \texttt{hallucination\_count} and \texttt{hallucination\_severity}.

\subsubsection{Readability Heuristic}
Readability was scored out of $100$ based on sentence length, complex word ratio, Flesch Reading Ease, and structural markers (lists, paragraphs). Each contributed up to $25$, summed as:

\[
Readability = \min(100, \sum_{k=1}^{4} Component_k)
\]

\subsubsection{Robustness and Fallback Logic}
Malformed JSON, missing fields, or model errors triggered full heuristic substitution. GPU memory fragmentation was mitigated by periodic cache clearing. Deterministic generation ($do\_sample=False$) ensured reproducibility, with $n\_runs>1$ enabling averaging.

\subsubsection{Threats to Validity}
Potential biases included upward score inflation due to clipped floors, false positives in hallucination detection, and limited semantic depth of lexical overlap. Despite these, the framework offered reproducible, multi-axis evaluation aligned with HCI priorities of safety, fairness, and inclusivity.

\begin{figure}[H]
    \centering
    \begin{subfigure}[t]{0.23\textwidth} 
        \centering
        \includegraphics[width=\linewidth]{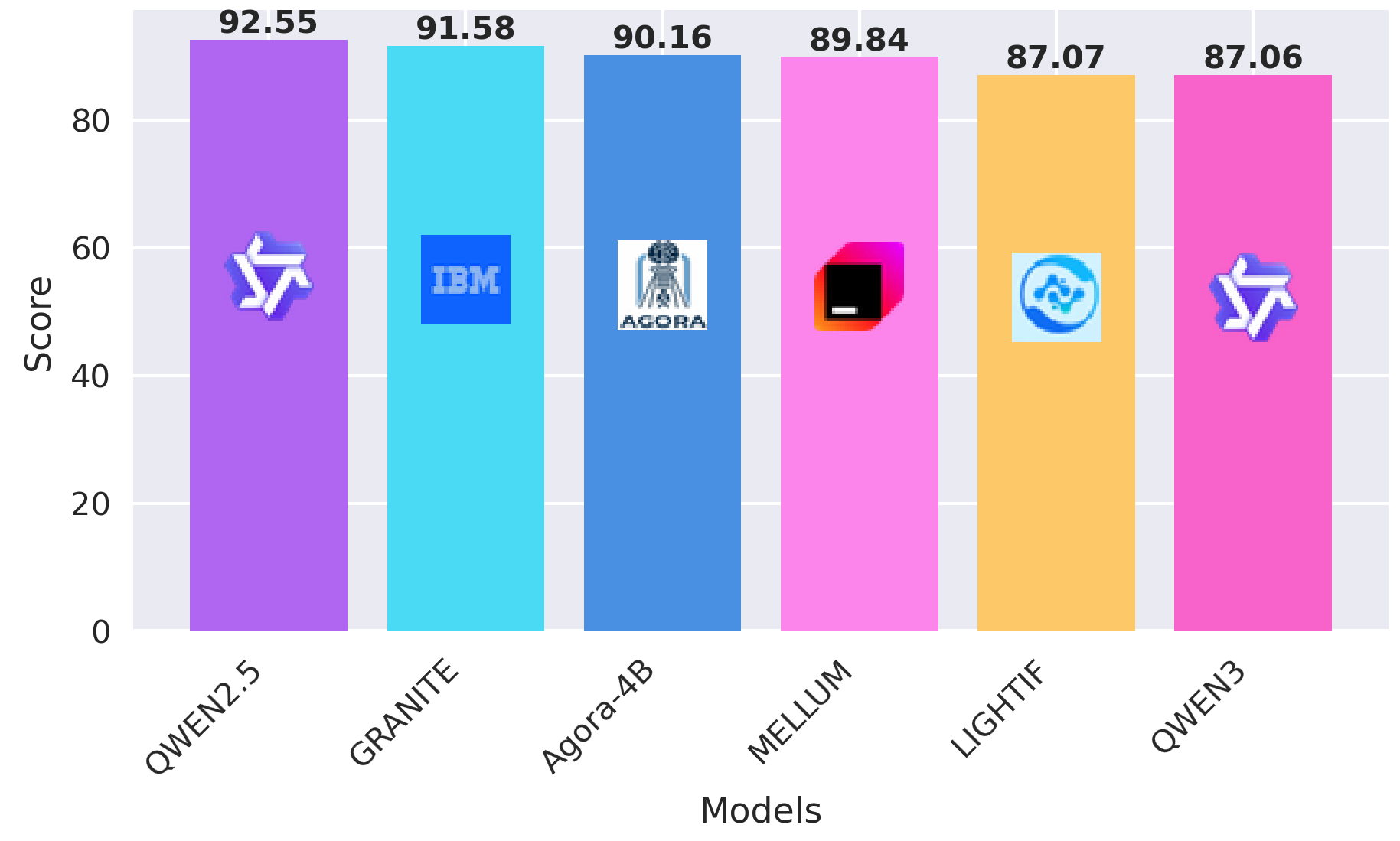}
        \caption{Concurrency Handling}
        \label{fig:7_1}
    \end{subfigure}
    \hfill
    \begin{subfigure}[t]{0.23\textwidth}
        \centering
        \includegraphics[width=\linewidth]{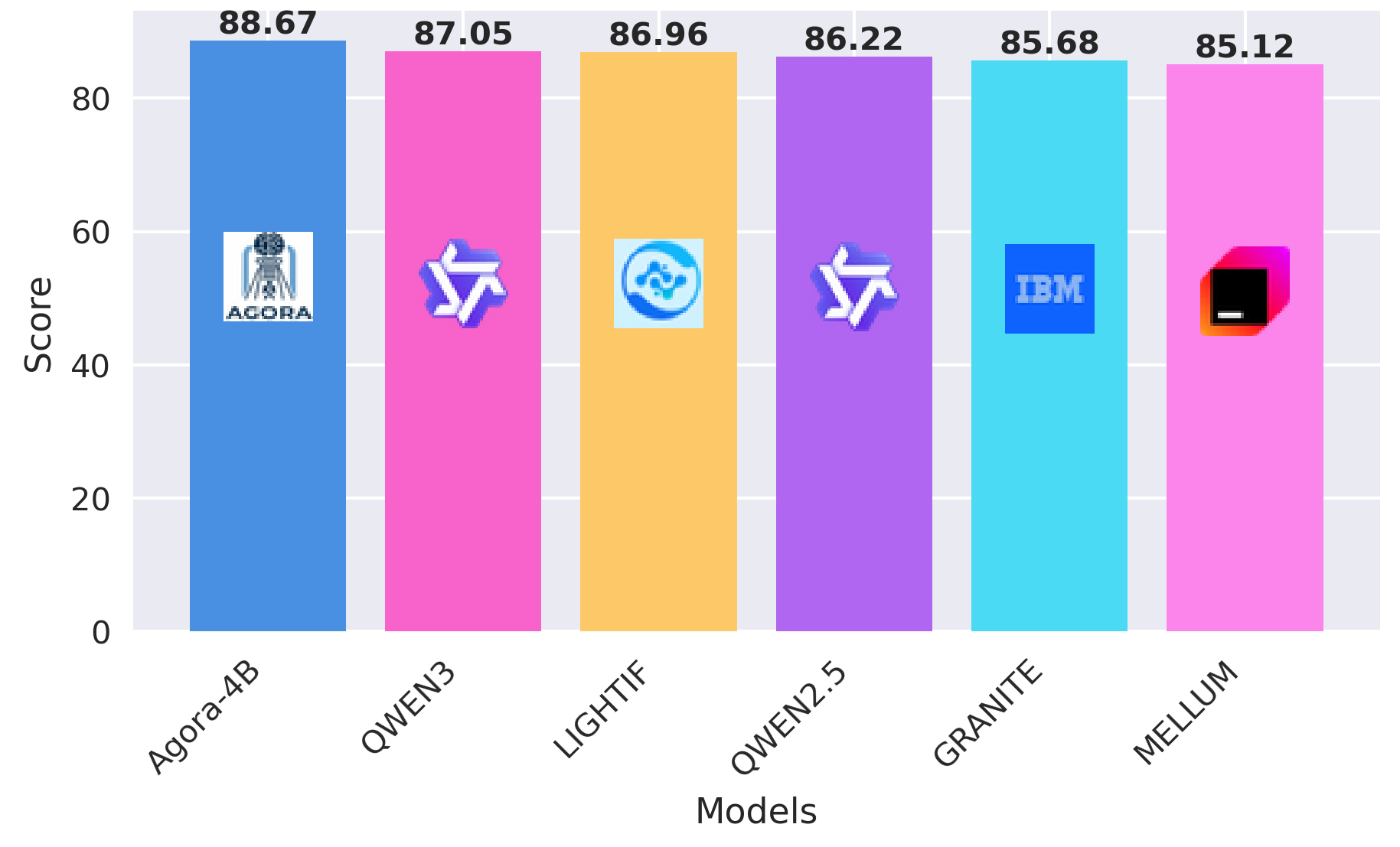}
        \caption{Relevance Coherence}
        \label{fig:7_2}
    \end{subfigure} 
    \hfill
    \begin{subfigure}[t]{0.23\textwidth}
        \centering
        \includegraphics[width=\linewidth]{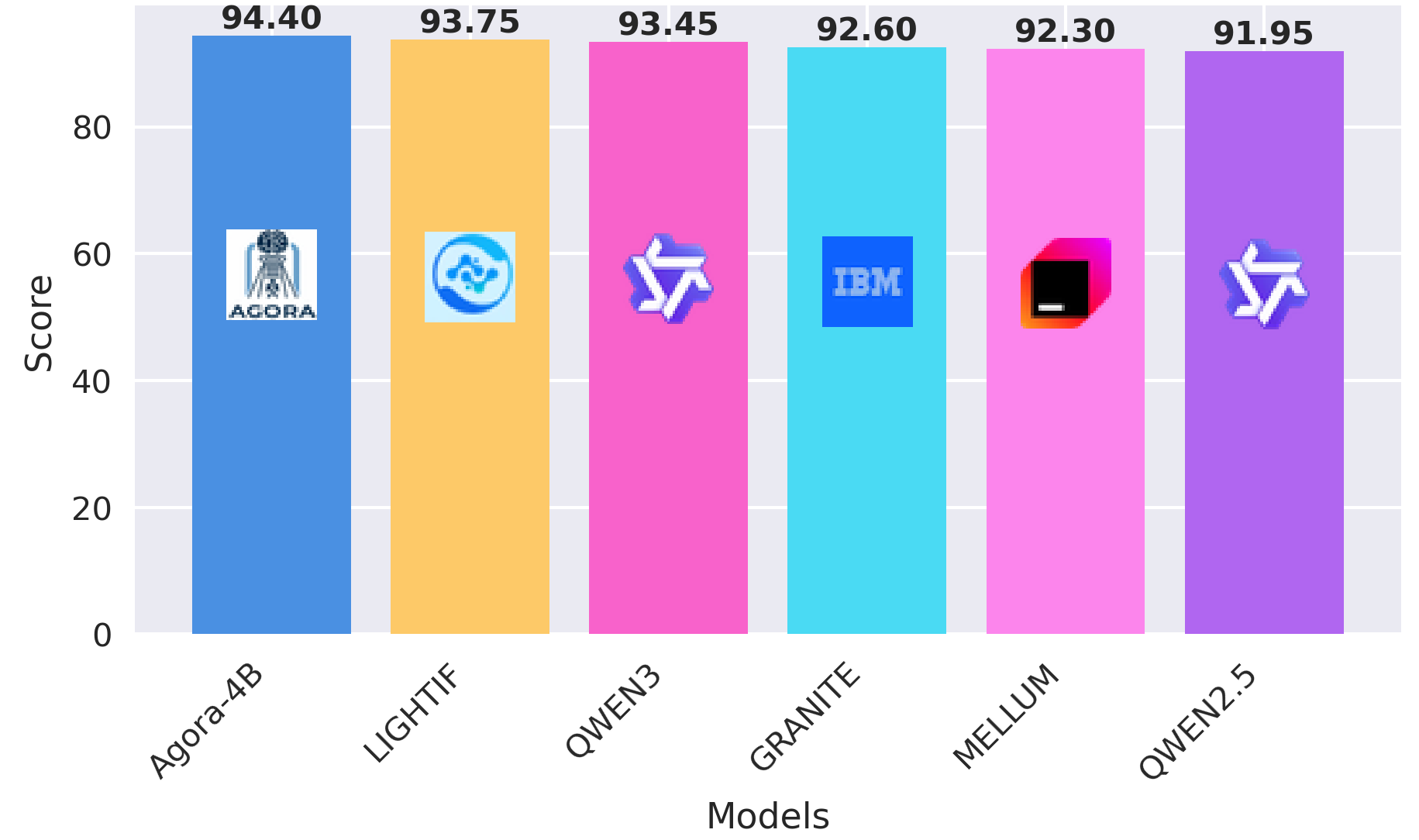}
        \caption{Linguistic Quality}
        \label{fig:7_3}
    \end{subfigure}
    \hfill
    \begin{subfigure}[t]{0.23\textwidth}
        \centering
        \includegraphics[width=\linewidth]{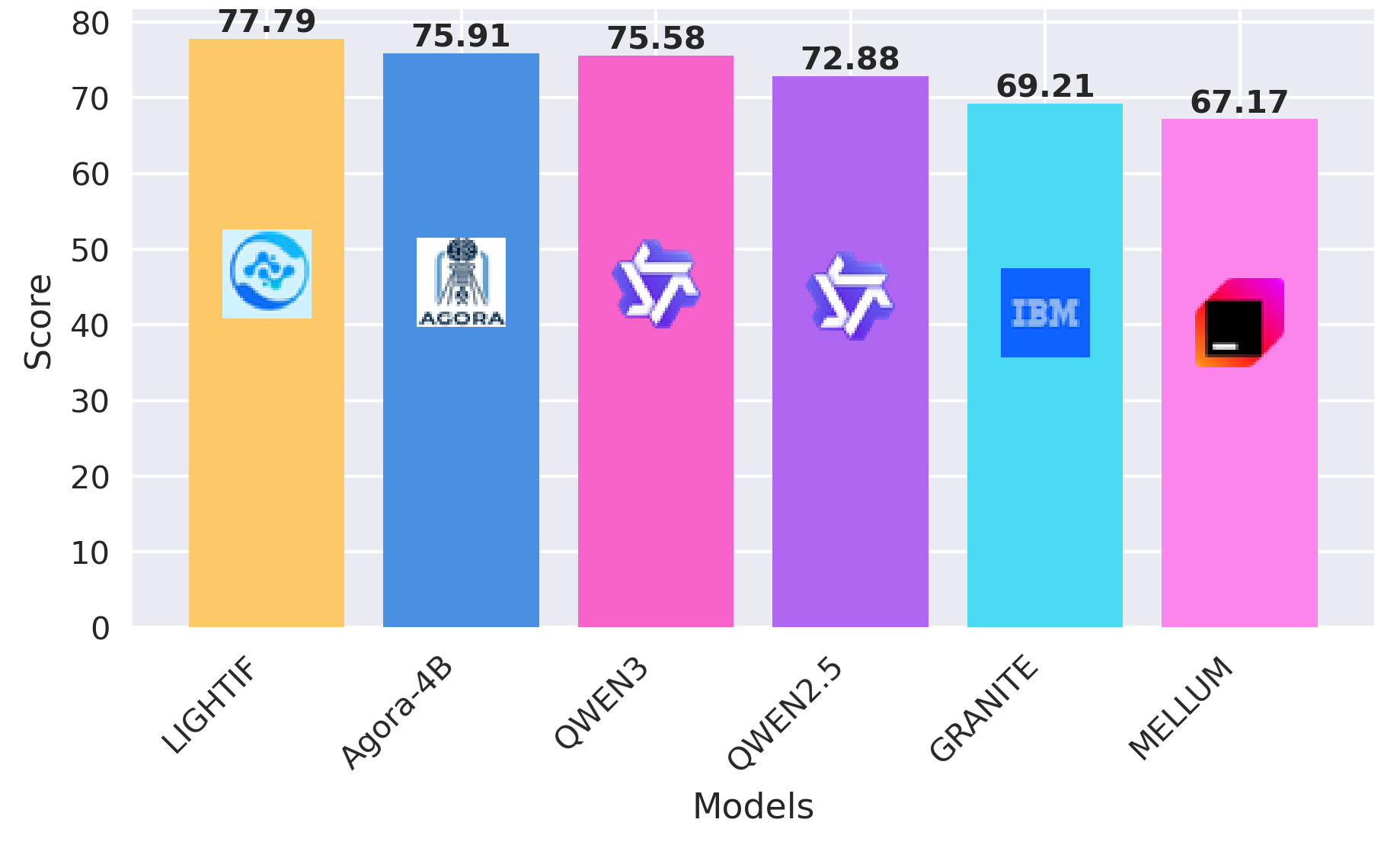}
        \caption{Response Accuracy}
        \label{fig:7_4}
    \end{subfigure}
    
    \vspace{1em} 
    
    \begin{subfigure}[t]{0.23\textwidth}
        \centering
        \includegraphics[width=\linewidth]{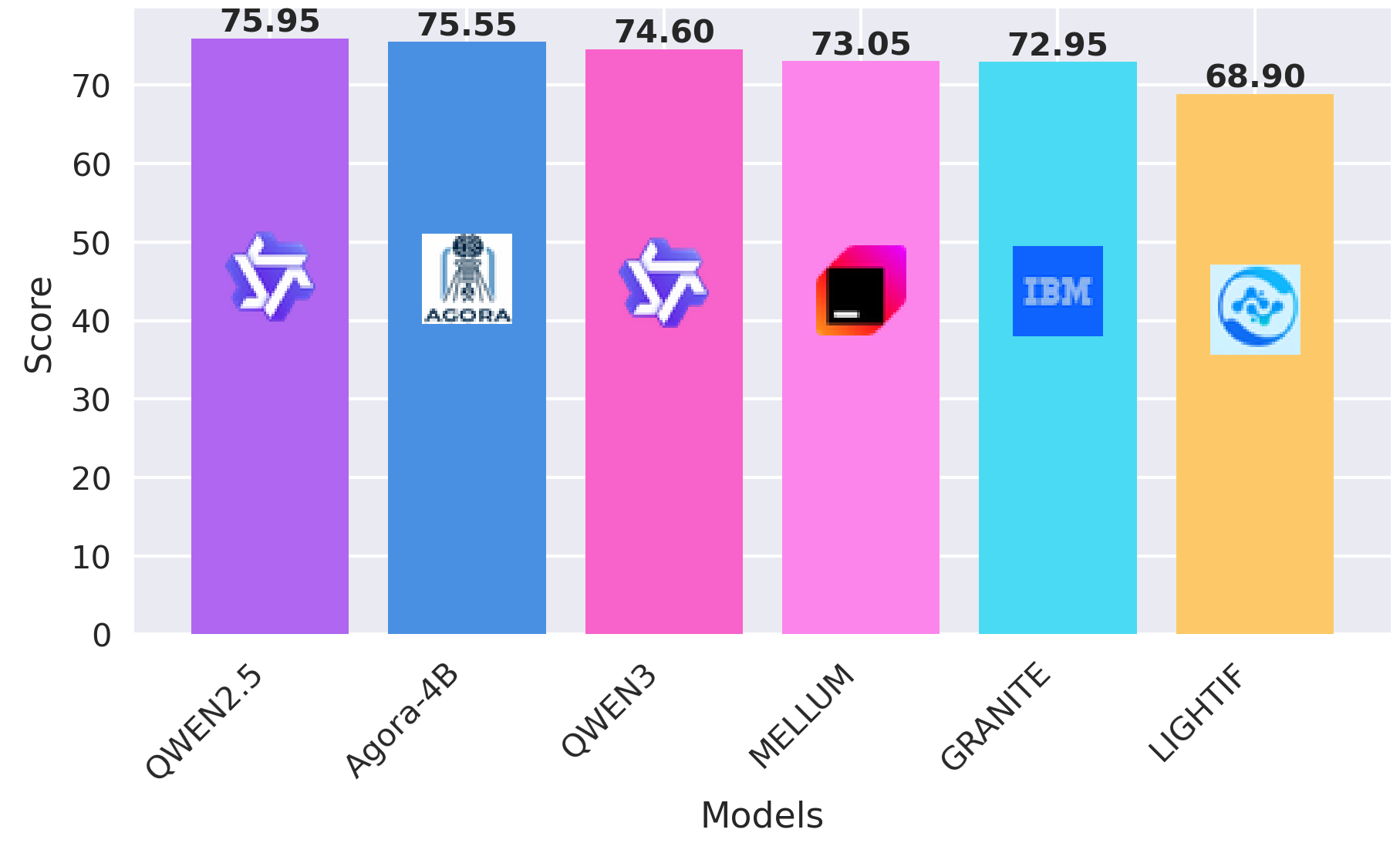}
        \caption{Hallucinations Detection}
        \label{fig:7_5}
    \end{subfigure} 
    \hfill
    \begin{subfigure}[t]{0.23\textwidth}
        \centering
        \includegraphics[width=\linewidth]{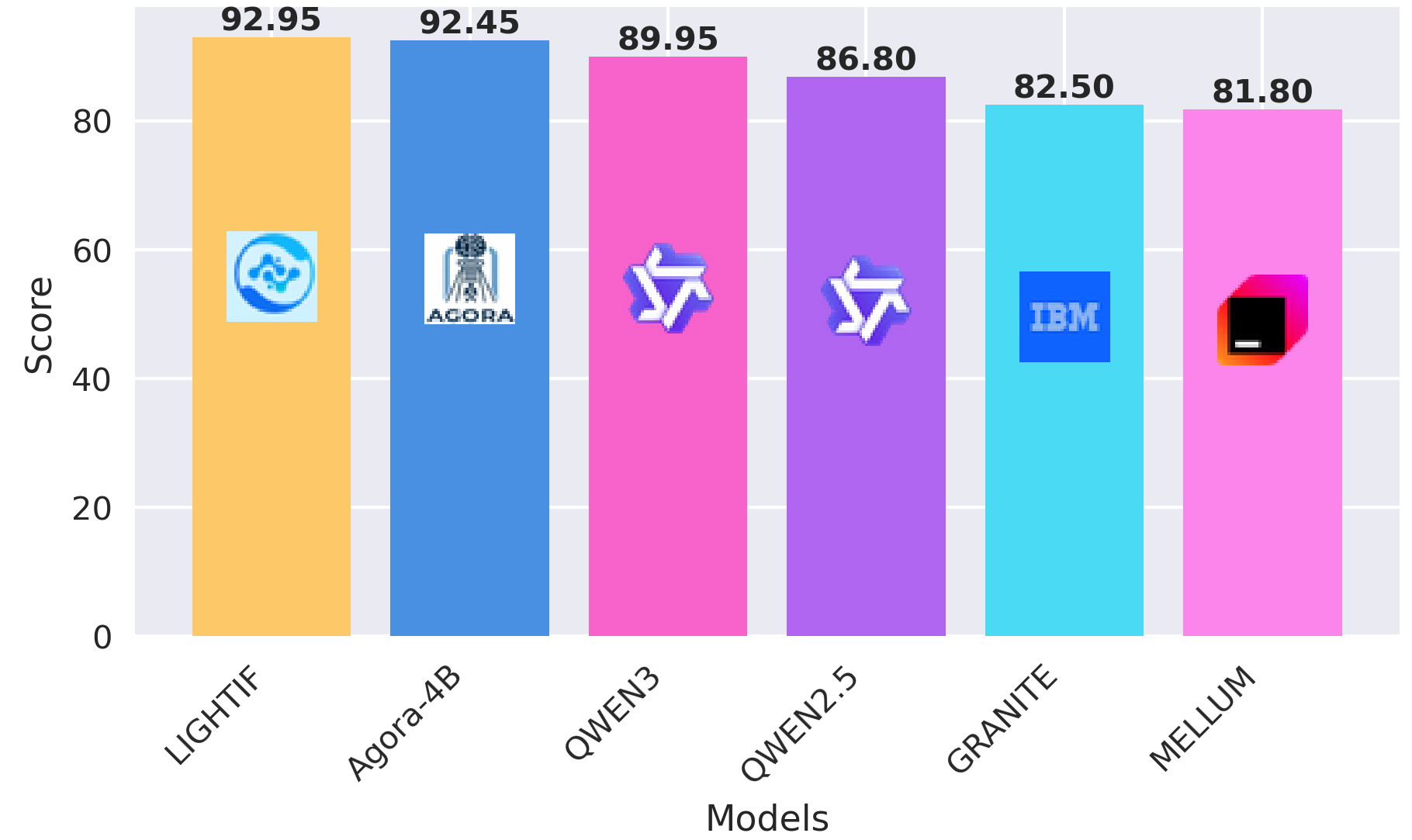}
        \caption{Trust and Safety}
        \label{fig:7_6}
    \end{subfigure}
    \hfill
    \begin{subfigure}[t]{0.23\textwidth}
        \centering
        \includegraphics[width=\linewidth]{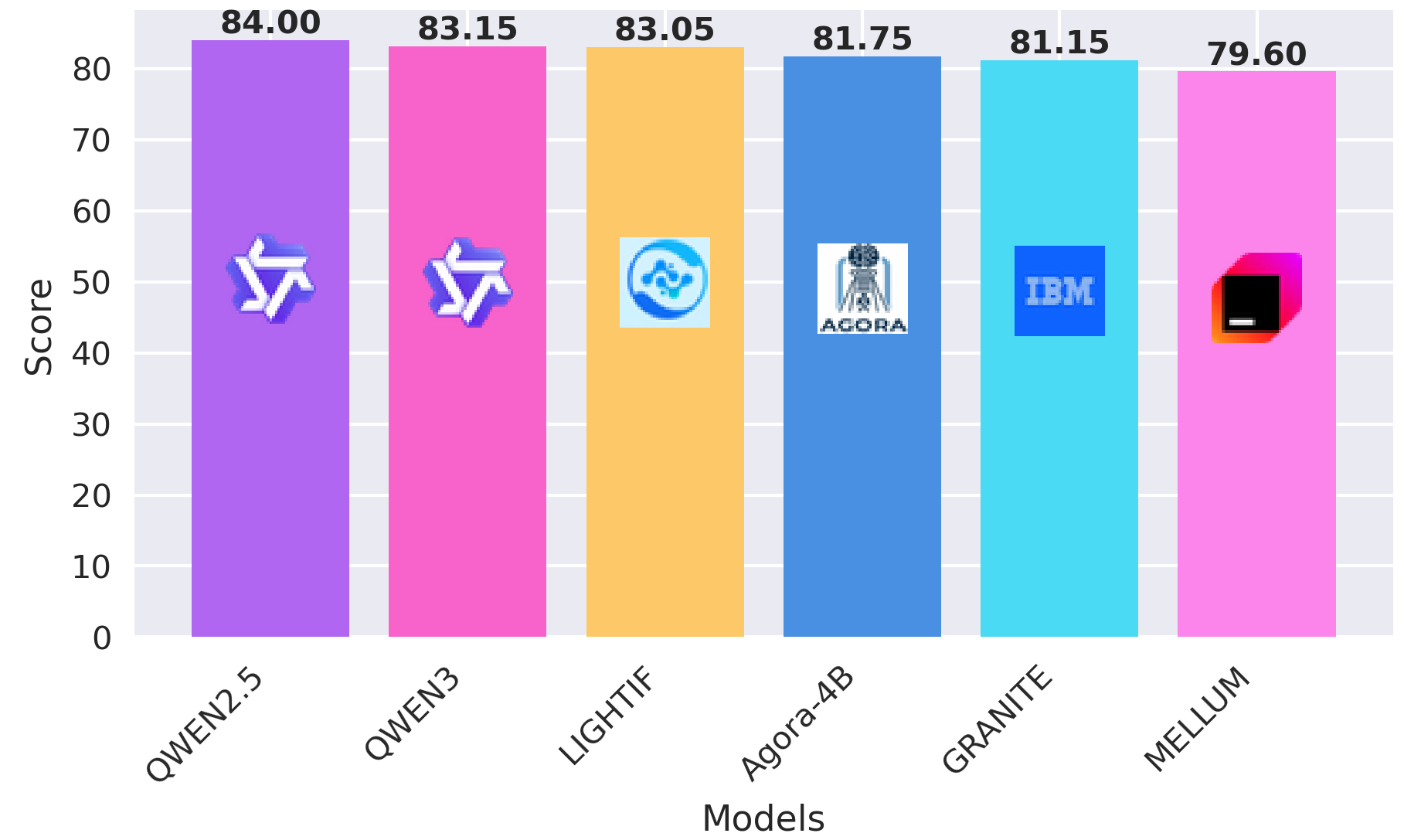}
        \caption{User Adaptation}
        \label{fig:7_7}
    \end{subfigure}
    \hfill
    \begin{subfigure}[t]{0.23\textwidth}
        \centering
        \includegraphics[width=\linewidth]{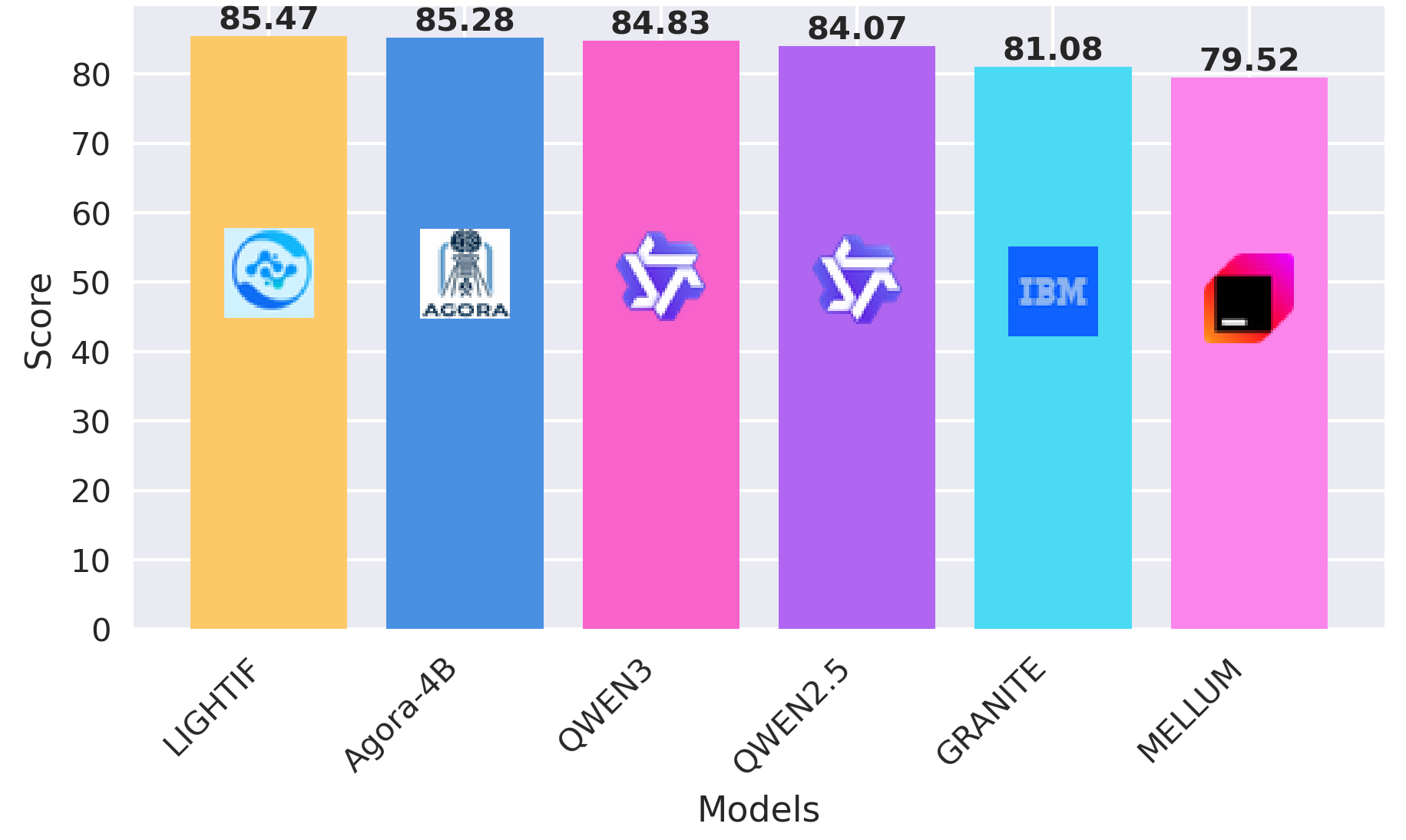}
        \caption{Total Score}
        \label{fig:7_8}
    \end{subfigure}
    
    \caption{Based on an extensive LLM-based evaluation, we present the top eight performance metrics for six leading models as they processed over 1,000 concurrent and individual queries. Our analysis shows that our model, \textbf{Agora 4B}, consistently ranks in the top three across every metric, a remarkable achievement given its significantly smaller size compared to its competitors.}
    \Description{Single-Model Performance Metrics Bar Charts
Eight vertical bar charts (a-h) comparing six single-agent models (Agora-4B in dark blue, QWEN2.5 in purple, GRANITE in pink, etc.) across metrics on 1,000+ queries, scaled 0-100. (a) Concurrency Handling: Agora-4B \~95, highest. (b) Relevance Coherence: Agora-4B \~89. (c) Linguistic Quality: LIGHTIF highest \~94, Agora-4B \~94. (d) Response Accuracy: Agora-4B \~82. (e) Hallucinations Detection: Agora-4B \~92. (f) Trust and Safety: Agora-4B \~92. (g) User Adaptation: QWEN2.5 \~84, Agora-4B \~82. (h) Total Score: Agora-4B \~85, top-ranked. The charts demonstrate Agora-4B's consistent top-three performance despite smaller size.}
    \label{fig:llm1}
\end{figure}

We tested our \textbf{Agora 4B} model's performance across 25+ metrics and top 8 metrics are displayed in the Fig. \ref{fig:llm1} compared to the leading LLMs in the same evaluation pipelines with same queries (both individual and concurrent were considered) and Agora-4B performs at Par and most of the times performs even better than models double of it's size. The detailed results can be see in Table. \ref{tab:model_performance_top5} with the top 6 metrics.

\begin{table}[htbp]
    \centering
    \caption{\textbf{Agora 4B}'s Performance vs Single Model Architecture using Leading Models}
    \Description{Agora-4B vs Single-Model Performance
A seven-column table comparing models. Columns: Model Name, Clarity and Tone, Linguistic Quality, Relevance and Coherence, User Adaptation, Trust and Safety, Total Score (all 0-100). Rows: Agora-4B (e.g., Clarity 96.40, Total 85.27), QWEN2.5 (84.06), GRANITE (81.08), QWEN3 (84.83), LIGHTIF (85.46), MELLUM (79.51). The table shows Agora-4B's competitive or superior scores in single-model setups}
    \label{tab:model_performance_top5}
    \resizebox{\textwidth}{!}{
\begin{tabular}{lrrrrrr}
\toprule
Model Name & Clarity and Tone & Linguistic Quality & Relevance and Coherence & User Adaptation & Trust and Safety & Total Score \\
\midrule
  \textbf{Agora-4B} & 96.40 & \textbf{94.40} & \textbf{88.67} & 81.75 & 92.45 & 85.27 \\
   QWEN2.5 & 98.25 & 91.95 & 86.22 & \textbf{84.00} & 86.80 & 84.06 \\
   GRANITE & 95.65 & 92.60 & 85.68 & 81.15 & 82.50 & 81.08 \\
     QWEN3 & 98.85 & 93.45 & 87.05 & 83.15 & 89.95 & 84.83 \\
   LIGHTIF & \textbf{99.30} & 93.75 & 86.96 & 83.05 & \textbf{92.95} & \textbf{85.46} \\
    MELLUM & 92.20 & 92.30 & 85.12 & 79.60 & 81.80 & 79.51 \\
\bottomrule
\end{tabular}
    }
\end{table}

\begin{figure}[H]
    \centering
    \begin{subfigure}[t]{0.23\textwidth} 
        \centering
        \includegraphics[width=\linewidth]{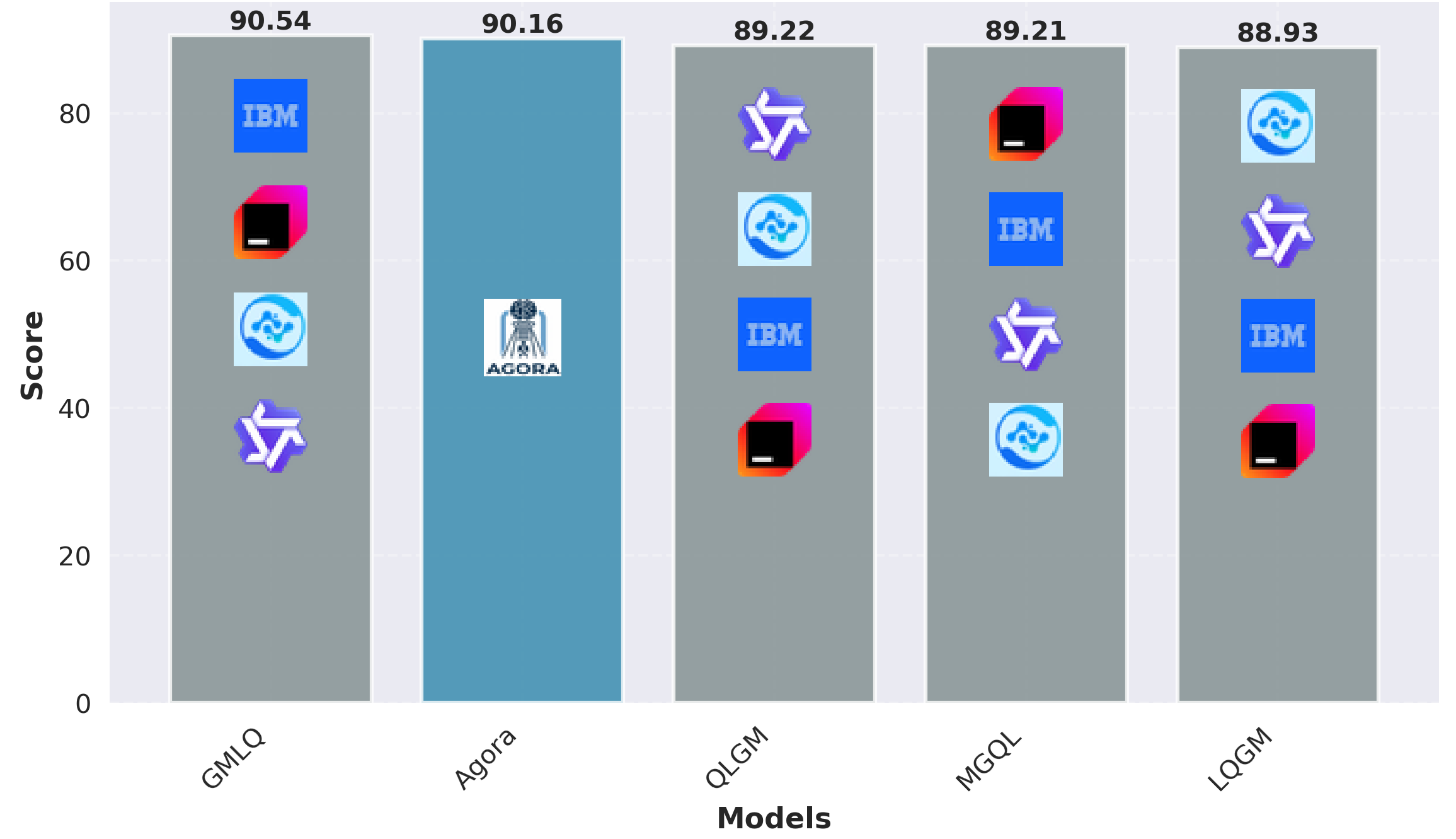}
        \caption{Concurrency Handling}
        \label{fig:7_1}
    \end{subfigure}
    \hfill
    \begin{subfigure}[t]{0.23\textwidth}
        \centering
        \includegraphics[width=\linewidth]{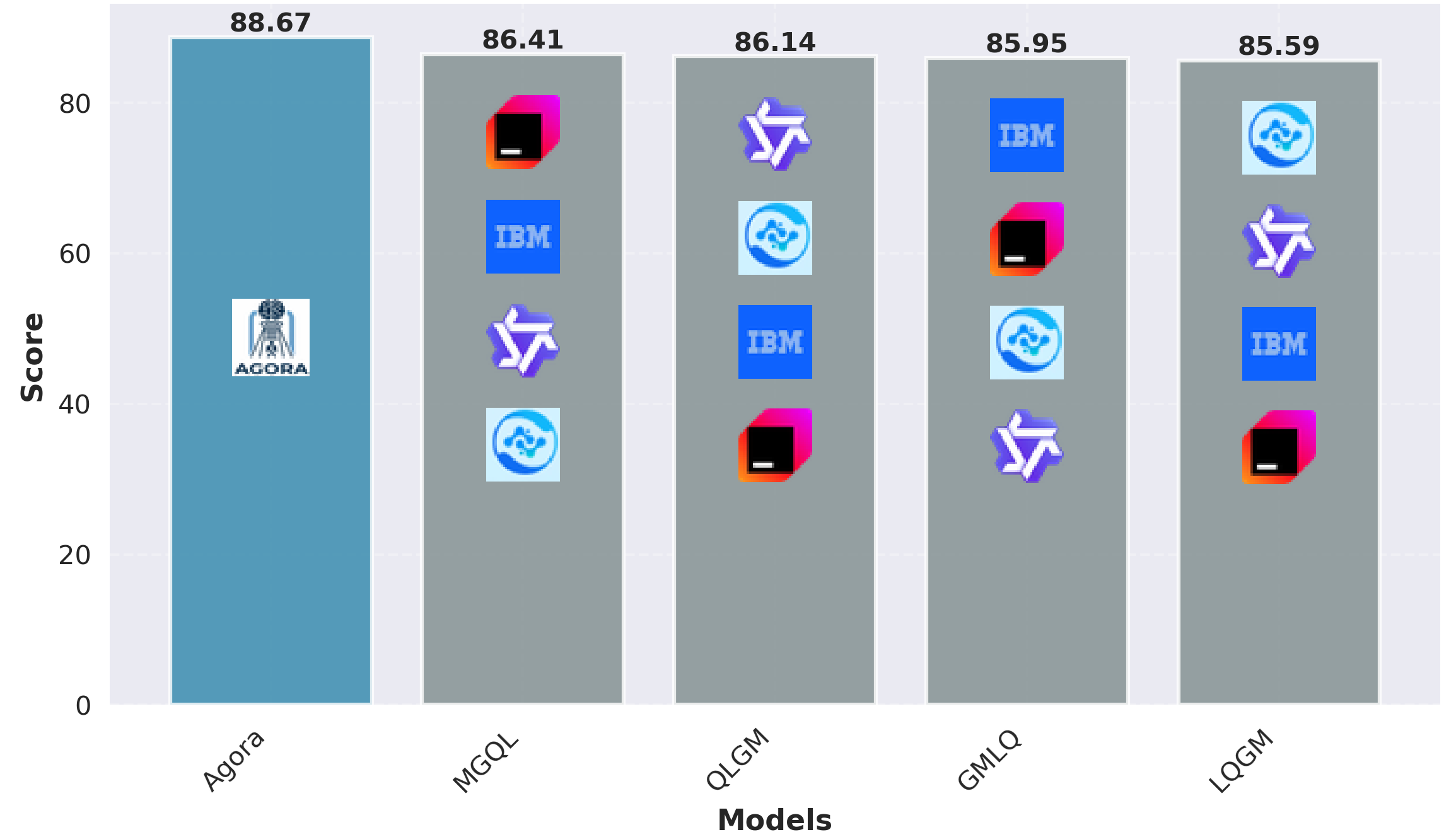}
        \caption{Relevance Coherence}
        \label{fig:7_2}
    \end{subfigure} 
    \hfill
    \begin{subfigure}[t]{0.23\textwidth}
        \centering
        \includegraphics[width=\linewidth]{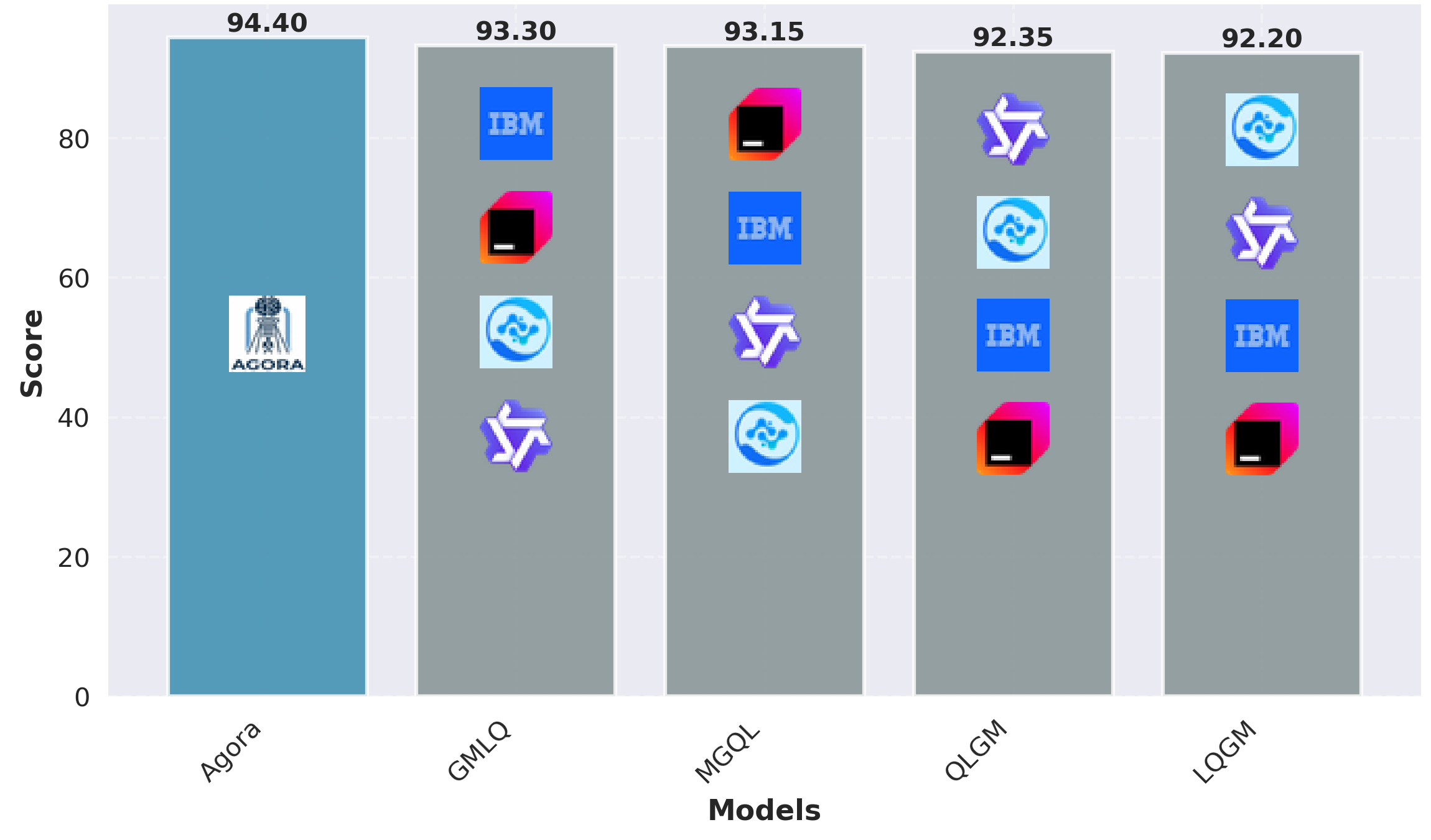}
        \caption{Linguistic Quality}
        \label{fig:7_3}
    \end{subfigure}
    \hfill
    \begin{subfigure}[t]{0.23\textwidth}
        \centering
        \includegraphics[width=\linewidth]{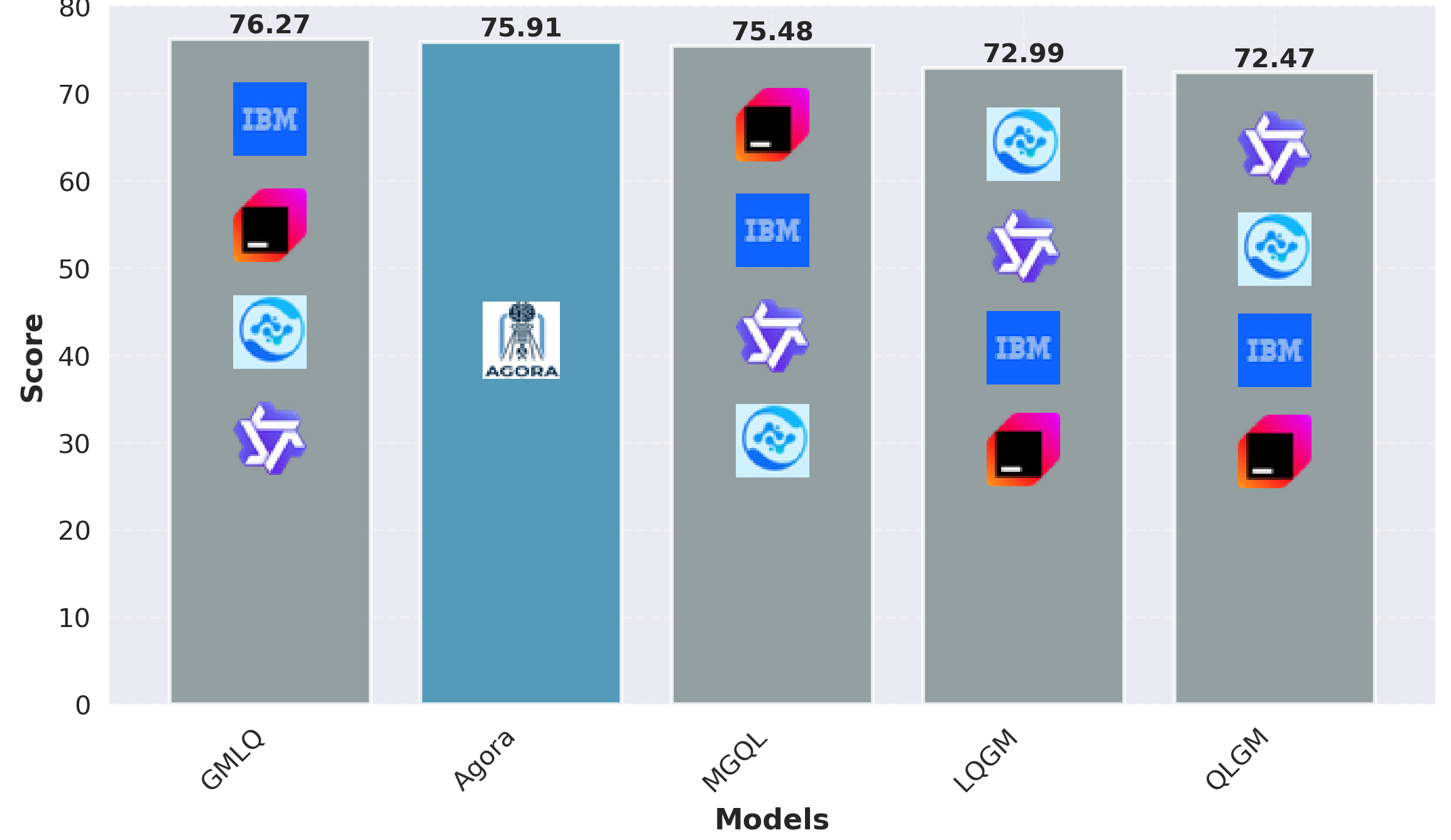}
        \caption{Response Accuracy}
        \label{fig:7_4}
    \end{subfigure}
    
    \vspace{1em} 
    
    \begin{subfigure}[t]{0.23\textwidth}
        \centering
        \includegraphics[width=\linewidth]{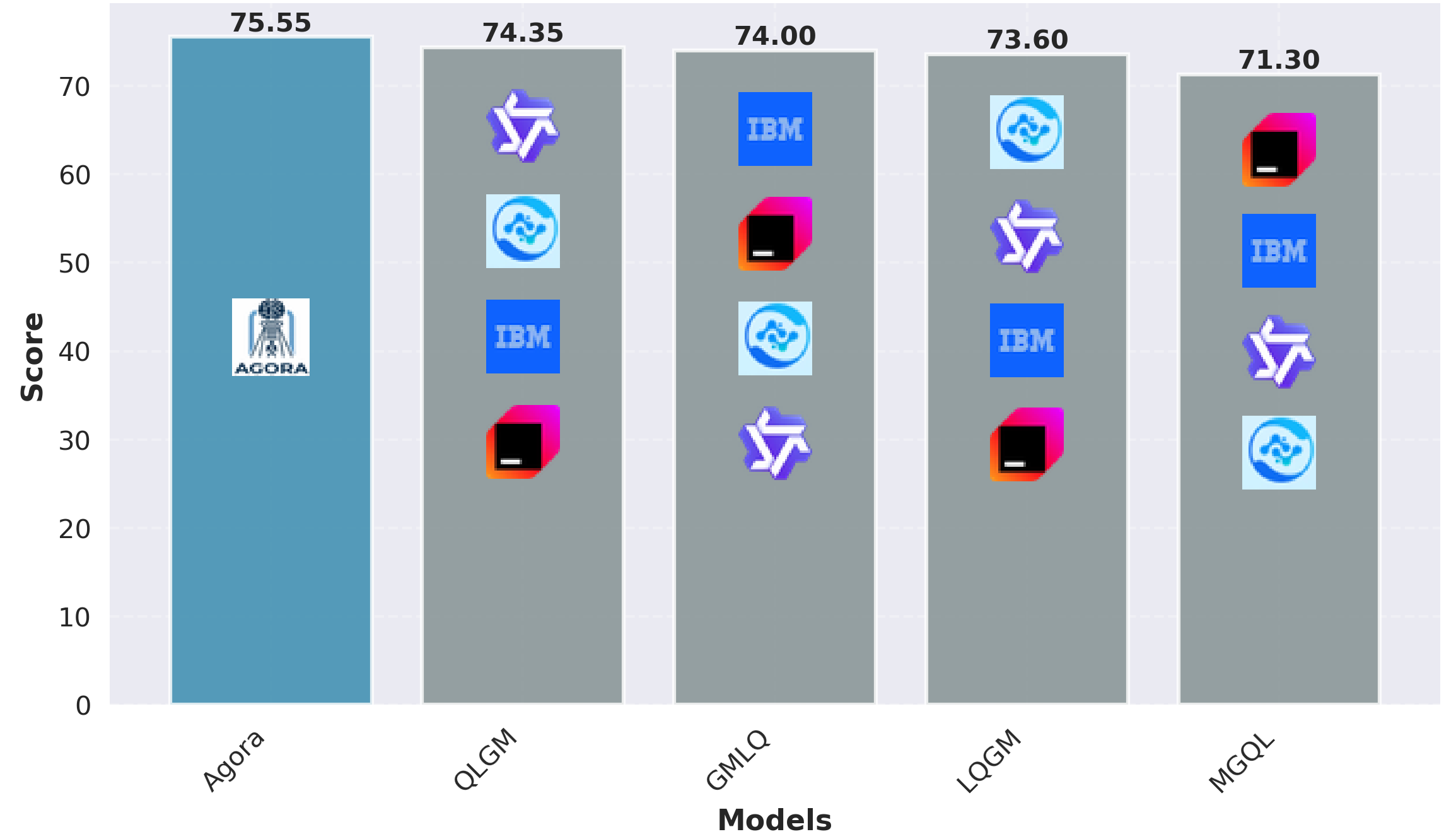}
        \caption{Hallucination Detection}
        \label{fig:7_5}
    \end{subfigure} 
    \hfill
    \begin{subfigure}[t]{0.23\textwidth}
        \centering
        \includegraphics[width=\linewidth]{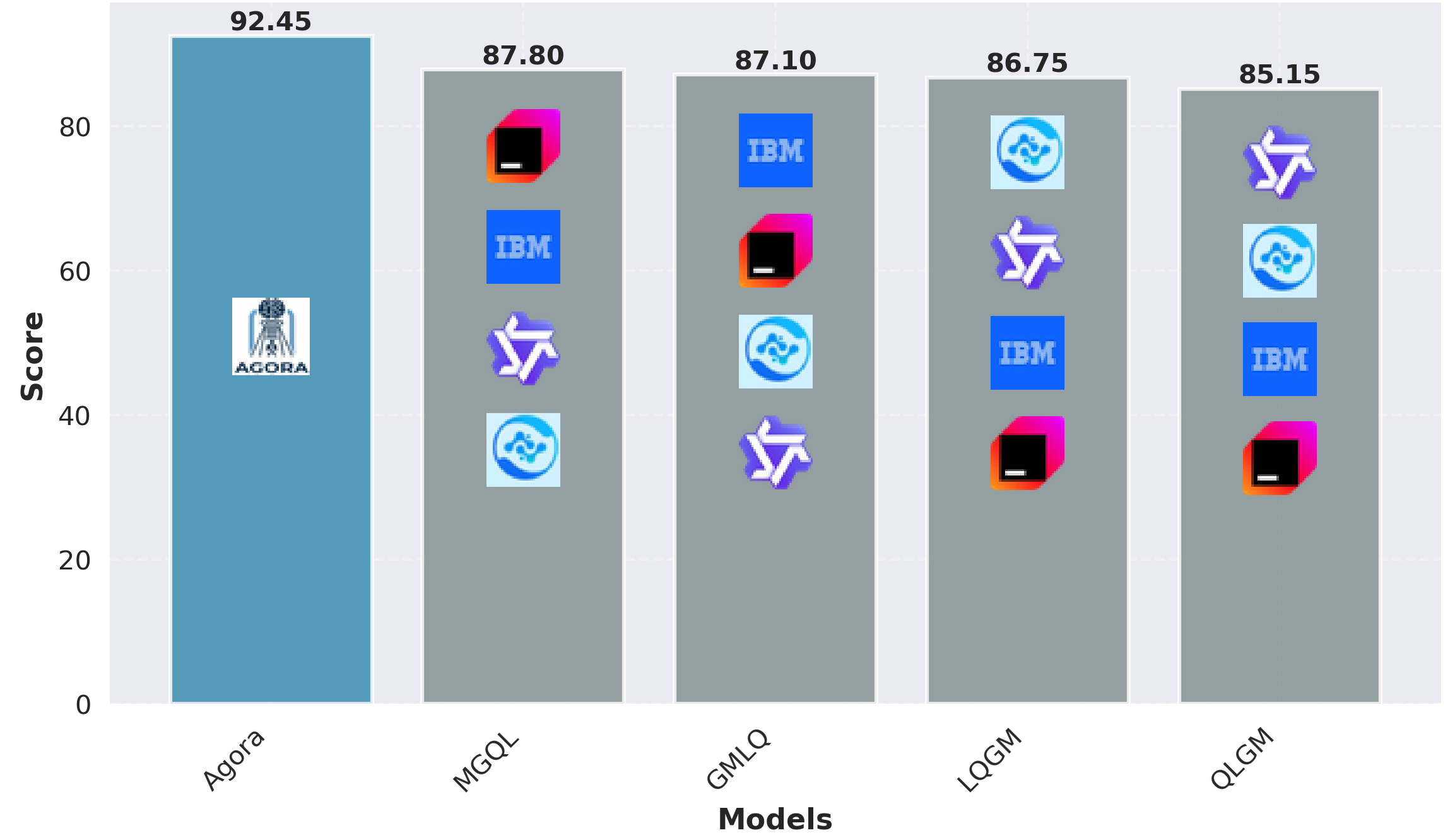}
        \caption{Trust and Safety}
        \label{fig:7_6}
    \end{subfigure}
    \hfill
    \begin{subfigure}[t]{0.23\textwidth}
        \centering
        \includegraphics[width=\linewidth]{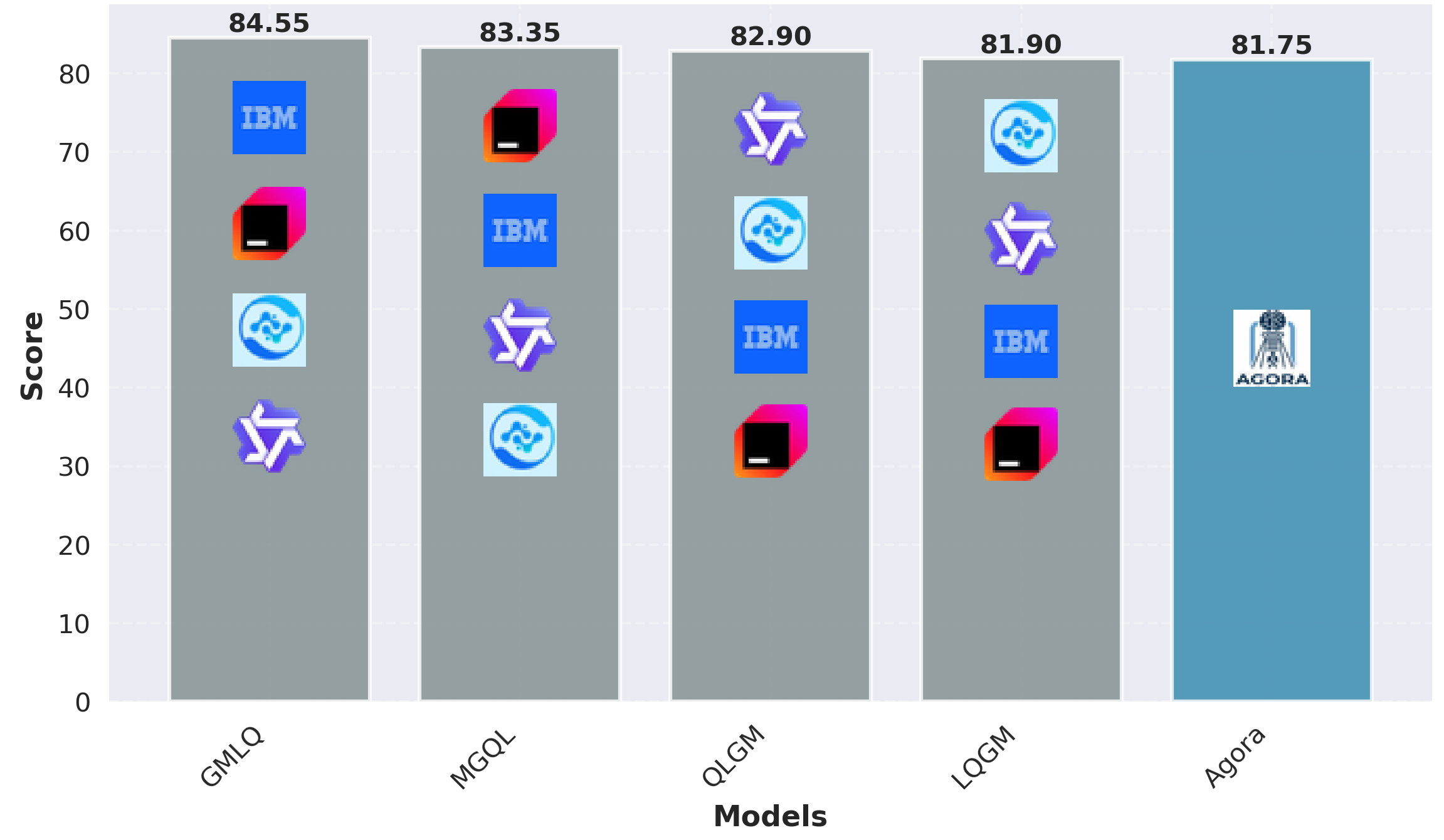}
        \caption{User Adaptation}
        \label{fig:7_7}
    \end{subfigure}
    \hfill
    \begin{subfigure}[t]{0.23\textwidth}
        \centering
        \includegraphics[width=\linewidth]{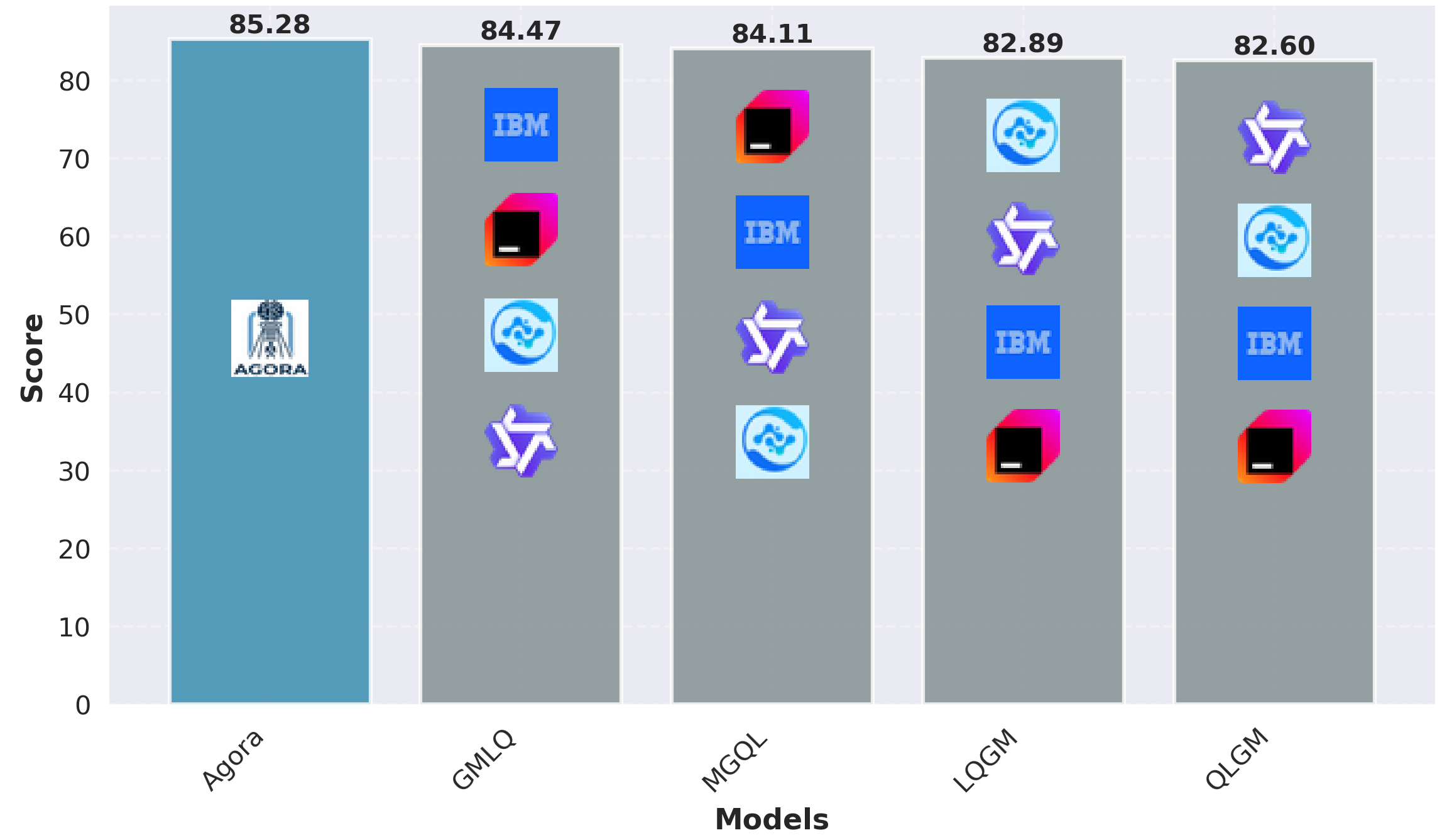}
        \caption{Total Score}
        \label{fig:7_8}
    \end{subfigure}
    
    \caption{Based on an extensive LLM-based evaluation, we present the top eight performance metrics for \textbf{Agora 4B} vs four leading multi-agent models, which were used to process over 1,000 concurrent and individual queries. Our analysis shows that our single-agent model, \textbf{Agora 4B}, consistently ranks in the top three across every metric. This is a remarkable achievement, given its significantly smaller size and its ability to outperform a system of four competing models in the multi agent system.}
    \Description{Single vs Multi-Agent Performance Metrics Bar Charts
Eight vertical bar charts (a-h) similar to Figure 9, but comparing Agora-4B (dark blue) against four multi-agent configurations (GMLQ in cyan, LQGM in green, etc.) on the same metrics. (a) Concurrency Handling: Agora-4B \~96, highest. (b) Relevance Coherence: Agora-4B \~89. (c) Linguistic Quality: GMLQ \~93, Agora-4B \~94. (d) Response Accuracy: Agora-4B \~82. (e) Hallucination Detection: Agora-4B \~92. (f) Trust and Safety: Agora-4B \~92. (g) User Adaptation: GMLQ \~85, Agora-4B \~82. (h) Total Score: Agora-4B \~85, outperforming multi-agents. The charts highlight the single-agent's superiority in efficiency and consistency.}
    \label{fig:llm2}
\end{figure}

We tested our single \textbf{Agora 4B} model based agent's performance across 25 metrics and top 8 metrics are displayed in the Fig. \ref{fig:llm2} compared to the leading multi LLMs in the same evaluation pipelines with same queries (both individual and concurrent were considered). Agora-4B performed better than most of the multi-agent systems by sufficient margins, where multiple models were used alternatively for different user types. In the Table. \ref{tab:model_performance_selected} top 6 metrics of testing can be seen, where Agora-4B outperforms most of the multi model agents. Here multi models were tested for different users iteratively for fair assessments and the names are written in abbreviations where G stands for IBM GRANITE Gurdian 3.2 5B \cite{padhi2024graniteguardian}, Q stands for Qwen2.5-7B model \cite{qwen2.5}, L stands for LIGHT-IF 8B model \cite{lightifproj} and M stands for Mellum 4B Base \cite{Mellum-4b-base}. Here are the brief descriptions of the evaluation parameters shown in \ref{fig:llm1} and \ref{fig:llm2} are:
\begin{itemize}
    \item \textbf{Concurrency Handling}: Scores multi-user interactions handling capabilities at the same (score: 0-100, weight: 0.08), ensuring equitable access.
    \item \textbf{Relevance Coherence}: Scores based on topical alignment and logical flow (0-100, weight: 0.05), enhancing user comprehension.
    \item \textbf{Linguistic Quality}: Scores based on evaluations grammar and readability (0-100, weight: 0.02), supporting accessibility.
    \item \textbf{Response Accuracy}: Measures semantic correctness and completeness (0-100, weight: 0.25), critical for task success.
    \item \textbf{Hallucination Detection}: Identifies fabricated content (0-100, weight: 0.10), promoting trust.
    \item \textbf{Trust and Safety}: Assesses ethical, harmless responses (0-100, weight: 0.20), prioritizing user well-being.
    \item \textbf{User Adaptation}: Scores based on responses tailored to diverse user types (0-100, weight: 0.15), ensuring inclusivity.
    \item \textbf{Total Score}: A weighted average (0-100), reflecting overall performance by taking the average of all the 25 parameters.
\end{itemize}

\begin{table}[htbp]
    \centering
    \caption{\textbf{Agora 4B}'s Performance vs Multi Model Architecture using Leading Models}
    \Description{Agora-4B vs Multi-Model Performance
A seven-column table similar to Table 3. Columns: Same metrics. Rows: Agora-4B (Total 85.28), GMLQ (84.47), LQGM (82.89), MGQL (84.11), QLGM (82.60). The table demonstrates Agora-4B outperforming multi-model agents in key metrics.}
    \label{tab:model_performance_selected}
    \resizebox{\textwidth}{!}{
\begin{tabular}{lrrrrrr}
\toprule
Model Name & Clarity and Tone & Linguistic Quality & Relevance and Coherence & User Adaptation & Trust and Safety & Total Score \\
\midrule
\textbf{Agora-4B} & 96.40 & \textbf{94.40} & \textbf{88.67} & 81.75 & \textbf{92.45} & \textbf{85.28} \\
GMLQ & 96.60 & 93.30 & 85.95 & \textbf{84.55} & 87.10 & 84.47 \\
LQGM & 96.10 & 92.20 & 85.59 & 81.90 & 86.75 & 82.89 \\
MGQL & \textbf{98.20} & 93.15 & 86.41 & 83.35 & 87.80 & 84.11 \\
QLGM & 95.25 & 92.35 & 86.14 & 82.90 & 85.15 & 82.60 \\
\bottomrule
\end{tabular}
    }
\end{table}

\subsection{Ethical Considerations and Researcher Positionality}
A strict ethical protocol was adopted for this study, especially in consideration of the sensitive nature of the data gathered. Informed consent was achieved for all participants, noting explicitly the purpose of the study, the character of the data gathered, and anonymization methods used. The application of GAI tools in gathering data was clearly stated, and the participants were be informed of possible risks involved, including the chance of "a malicious agent can break into and extract chat histories". To ensure that this is avoided, there will be a strong protocol for anonymizing data and security before being used in the APP. All people participating in this survey were assigned nicknames in order to hide their real names and don't fall victim to any possible data leakage by the agent.
Key to this, the study includes an exploration of researcher positionality, recognizing that the researchers' own Neurodivergent selves and experiences with GAI will inform the data collection and analysis. The autoethnographic process, as exemplified by the research model presented here, acts as a model of self-reflection and openness to ensure that researchers' own backgrounds and biases are seen as inherent to the research process. This process of critical self-reflection is essential to ensuring the integrity and validity of the findings.

Designing AI systems that truly work for everyone, especially vulnerable groups like the elderly, children, and Neurodivergent individuals, means involving them in the process from the start. Participatory design (PD) has long been championed in human-computer interaction (HCI) for bringing users into “collective decision-making.” \cite{10.1145/3708557.3716334} Historically, vulnerable groups have been left out, and studies show that failing to include them results in systems that don’t meet their needs or even cause harm. \cite{cesaroni2025participatorystrategyaiethics} For example, involving older adults in co-design helps figure out what information they need and how to present it, leading to better smart-home interfaces and greater adoption. \cite{Chan2024co} Similarly, research on kids and AI stresses giving them a voice, noting that while AI is everywhere in children’s lives, “children’s involvement in the development of these systems appears to be minimal… and their own views… are, crucially, missing.” \cite{voku} To address this, accessible PD methods like craft-based ideation, storytelling, and role-play—have worked well for engaging Neurodivergent and young users. \cite{SHARMA2022100521} Recent studies on accessible PD for kids highlight the value of multi-modal approaches that support different learning and communication styles. \cite{10.1145/3613904.3642283} Our approach builds on this by using multi-sensory co-design tools in workshops tailored to diverse users (see Methodology).

Beyond HCI, AI ethics research points out gaps between technological advances and their impact on society. Scholars push for weaving ethical values like privacy, fairness, and transparency into the entire AI development process. \cite{CORREA2023100857} Human-Centered AI (HCAI) frameworks call for ongoing user involvement to ensure systems reflect human values. \cite{shneiderman2020humancenteredartificialintelligencereliable} For instance, the IEEE Ethics Guidelines and other standards emphasize including stakeholders, especially vulnerable users, from the beginning to tackle biases and trust issues. \cite{PS2023100165} Pilot studies with AI professionals also highlight concerns about bias, privacy, and accessibility, stressing “the need for greater awareness and structured dialogue to integrate the IDEAS principles throughout the AI lifecycle.” \cite{sci6010003} Our work builds on this by focusing on home settings and drawing practical insights from user-driven data.
Smart home technology is a mixed bag, it can improve life with things like energy savings or health monitoring, but it also brings risks. A recent review found that smart home research often focuses on tech but overlooks “user-centric” perspectives on harms. \cite{10.1145/3706598.3713494} Privacy breaches, security weaknesses, and well-being concerns frequently pop up in risk analyses. \cite{haron} That review also noted a key gap: more work is needed on building user agency and autonomy into smart-home design. \cite{10.1145/3476084} We address this by exploring conversations that show how much users value control, consent, and clear interfaces in home AI through the Autonomy Slider.

\section{Limitations \& Future Work}
\subsection{Limitations}

Several limitations constrain the scope and generalizability of this work \cite{Ceuterick2020, 10.1145/3638380.3638412}. First, the evaluation relied mostly on synthetic and curated datasets as limited amount of real human data was available in homely set up, which, while carefully constructed, may sanitize the complexity of real domestic interactions \cite{DU2023e21448, PANDE2024102580}. Nuances such as overlapping speech, cultural variation, and longitudinal shifts in household dynamics were not fully captured. Although the synthetic curriculum introduced fairness and safety constraints, this process may have inadvertently encoded biases from the base models, raising the risk that inclusivity is overstated.

Second, the empirical study involved a relatively small and demographically narrow sample of participants. While efforts were made to recruit children, older adults, and Neurodivergent individuals, the study did not capture the diversity of socioeconomic, linguistic, or cultural contexts in which domestic AI systems will operate. Consequently, the findings should be read as indicative rather than definitive.

Third, the evaluation pipeline itself may reinforce the very issues it aims to surface. Heuristic fallback scoring was intentionally generous, potentially masking critical system failures. Similarly, hallucination detection based on regex patterns risks both false positives (flagging legitimate content) and false negatives (overlooking subtle misalignment). This raises questions about the robustness of the reported quantitative results.

Finally, the prototype was assessed in controlled settings in deployments within real households. The dynamics of everyday domestic life, including interruptions, conflicts, and gradual trust calibration remain a very small portion of the survey as it's not possible or desirable to arise that kind of issues in home every time. As such, the system’s long-term acceptance, ethical implications, and failure modes in the wild are yet to be understood. These limitations highlight the need for cautious interpretation and future work that engages more deeply with lived contexts, diverse communities, and longitudinal evaluation.

\subsection{Future Work}
Future research should extend these findings in several directions \cite{10.1145/3715275.3732136, boyd01062012}. First, larger-scale and longitudinal deployments in diverse households would provide insights into sustained trust, adaptation, and fairness across time \cite{Chang2024, 10.1109/MC.2013.202}. Second, integrating semantic similarity methods, such as embedding-based alignment, could reduce the limitations of lexical overlap in response evaluation and improve hallucination detection beyond rule-based heuristics \cite{zhang2023sirenssongaiocean, Zhang_2022}. Third, expanding the IDEAS framework into continuous monitoring tools could enhance proactive bias audits and safety interventions in real-world use. Fourth, future prototypes may incorporate richer multi-modal sensing (e.g., emotion recognition, ambient context signals) to support deeper personalization for children, older adults, and Neurodivergent users \cite{GAO2017410, 10.1145/3173574.3174214, Ma2022}. Fifth, co-design should be broadened across cultural contexts to validate inclusivity principles globally, especially in households with differing caregiving structures \cite{Chung2016, Edwards2016, Alfeir2024}. Finally, longitudinal studies on multi-agent conflict resolution and adaptive autonomy sliders could reveal how inclusive AI systems mediate tensions in plural domestic spaces over time \cite{wen2025families, 10.1145/3025453.3025739, Elish2019}.

\section{Conclusion}
This work introduced the Plural Voices Model (PVM), a single-agent framework designed to negotiate the competing needs of diverse household members through value-sensitive adaptation, privacy-preserving deployment, and inclusive design. By curating and extending open datasets with a synthetic curriculum, and embedding ethical scaffolds into the Agora-4B foundation, we demonstrated how domestic AI can balance autonomy, safety, and fairness across children, older adults, Neurodivergent users, and typical adults. Comparative evaluations showed that PVM outperformed larger multi-agent baselines on compliance, fairness, and safety while remaining computationally efficient. Beyond technical performance, this research contributes design principles for inclusive domestic AI, such as participatory co-design, transparency mechanisms, adaptive interaction modalities, and family-centered governance features. Together, these findings advance a vision of agentic AI as a collaborative partner rather than an opaque controller in everyday life. At the same time, significant challenges remain. Synthetic datasets risk oversimplifying lived experiences; small-scale, short-term studies cannot capture the complexities of long term longitudinal household use; and LLM based evaluation pipelines may inadvertently reproduce the very biases they aim to measure. Addressing these tensions requires deeper integration of participatory methods, richer ecological deployments, and broader cross-cultural validation. By foregrounding vulnerable voices and emphasizing ethical, transparent, and user-centered design, this work aims to chart a path for future domestic AI that not only responds to commands but also respects values, identities, and plural household dynamics.

\section*{Declaration of generative AI and AI-assisted technologies in the writing process}
During the preparation of this work the authors used ChatGPT, Gemini and Grok in order to cut down repetitions and improve readability and language. After using this tool/service, the authors reviewed and edited the content as needed and takes full responsibility for the content of the publication.

\bibliographystyle{plain}
\bibliography{references}

\clearpage
\label{TotPages}
\end{document}